\documentclass[preprint2,numberedappendix]{aastex6}

\usepackage{natbib}
\usepackage{multirow}
\usepackage{gensymb} 
\usepackage[flushleft]{threeparttable} 
\usepackage{amsmath}
\usepackage{url}

\newcommand{\Ks}{$K_\mathrm{s}$ }

\shorttitle{Rest-frame $z'$  LF at $4\lesssim z \lesssim 7$}
\shortauthors{Stefanon et al.}

\begin{document}

\title{The Rest-Frame Optical (900\MakeLowercase{nm}) Galaxy Luminosity Function at \MakeLowercase{z}$\sim4-7$: Abundance Matching Points to Limited Evolution in the $M_\mathrm{star}/M_\mathrm{halo}$ ratio  at \MakeLowercase{z}$\ge4$}

\author{Mauro Stefanon\altaffilmark{1}, Rychard J. Bouwens\altaffilmark{1}, Ivo Labb\'e\altaffilmark{1}, Adam Muzzin\altaffilmark{2}, Danilo Marchesini\altaffilmark{3}, Pascal Oesch\altaffilmark{4}, Valentino Gonzalez\altaffilmark{5}}

\email{Email: stefanon@strw.leidenuniv.nl}

\altaffiltext{1}{Leiden Observatory, Leiden University, NL-2300 RA Leiden, Netherlands}
\altaffiltext{2}{Department of Physics and Astronomy, York University, 4700 Keele St., Toronto, Ontario, Canada, MJ3 1P3}
\altaffiltext{3}{Physics and Astronomy Department, Tufts University, Robinson Hall, Room 257, Medford, MA, 02155, USA}
\altaffiltext{4}{Observatoire de Gen\`eve, 1290 Versoix, Switzerland}
\altaffiltext{5}{Departamento de Astronomia, Universidad de Chile, Santiago, Chile}

\begin{abstract}

We present the first determination of the galaxy luminosity function (LF) at $z\sim 4, 5, 6$ and $7$ in the rest-frame optical at $\lambda_\mathrm{rest}\sim900$~nm ($z'$ band).  The rest-frame optical light traces the content in low-mass evolved stars ($\sim$stellar mass - $M_*$), minimizing potential measurement biases for  $M_*$. Moreover it is less affected by  nebular line emission contamination and  dust attenuation,  is independent of stellar population models, and can be probed up to $z\sim8$ through {\it Spitzer}/IRAC. Our analysis leverages the unique full depth \emph{Spitzer}/IRAC $3.6\mu$m-to-$8.0\mu$m data over the CANDELS/GOODS-N, CANDELS/GOODS-S and COSMOS/UltraVISTA fields. We find that at absolute magnitudes $M_{z'}$ fainter than $\gtrsim-23$~mag, $M_{z'}$ linearly correlates with $M_\mathrm{UV,1600}$. At brighter $M_{z'}$,  $M_\mathrm{UV,1600}$ presents a turnover, suggesting that the stellar mass-to-light ratio $M_*/L_\mathrm{UV,1600}$ could be characterised by a very broad range of values at high stellar masses.  Median-stacking {analyses recover} a $M_*/L_{z'}$ roughly independent on $M_{z'}$ for $M_{z'}\gtrsim-23$~mag, but exponentially increasing at brighter magnitudes. We find that the evolution of the LF marginally prefers a pure luminosity evolution over a pure density evolution, with the characteristic luminosity decreasing by a factor $\sim5\times$ between $z\sim4$ and $z\sim7$. Direct application of the recovered $M_*/L_{z'}$ generates stellar mass functions consistent with average measurements from the literature. Measurements of the stellar-to-halo mass ratio at fixed cumulative number density show that it is roughly constant with redshift for $M_h\gtrsim10^{12}M_\odot$. This is also supported by the fact that the evolution of the LF at $4 \lesssim z \lesssim 7$ can be accounted for by a rigid displacement in luminosity corresponding to the evolution of the halo mass from abundance matching.

\end{abstract}


\section{Introduction}

The stellar mass is one of the most fundamental parameters characterising galaxies. This observable is driven by the light emitted in the rest-frame optical/near infrared (NIR) by lower mass stars and it correlates with the dynamical mass up to $z\sim2$ (e.g., \citealt{cappellari2009,taylor2010,vandesande2014,barro2014}), suggesting that it can be a robust estimate of the cumulative content of matter in galaxies. Stellar masses have been estimated for galaxies at redshifts as high as $z\sim7-8$ (e.g., \citealt{labbe2010a, labbe2013}). Moreover, stellar mass estimates are readily available in the models of galaxy formation and evolution. For the above reasons, the stellar mass has been largely adopted in comparisons to the models. 

The stellar mass function (SMF), i.e., the number density of galaxies per unit (log) stellar mass, provides a first census of a galaxy population and it is therefore  one of the most basic observables that need to be reproduced by any successful model of galaxy formation. Multi-wavelength photometric surveys like UltraVISTA \citep{mccracken2012} and ZFOURGE  \citep{straatman2016} have enabled SMF measurements up to $z\sim3-4$ (e.g., \citealt{ilbert2013,muzzin2013b,tomczak2014}, but see also \citealt{stefanon2015, caputi2015} for SMF of massive galaxies up to $z\sim7$), whereas {\it HST} surveys, like CANDELS (\citealt{koekemoer2011,grogin2011}), GOODS (\citealt{giavalisco2004}) and HUDF (\citealt{beckwith2006, bouwens2011}), complemented by {\it Spitzer}/IRAC observations (e.g., Spitzer/GOODS - PI Dickinson; \citealt{ashby2015}),  extended the study to the evolution of the SMF up to $z\sim7$ (e.g., \citealt{stark2009,gonzalez2011,duncan2014,grazian2015,song2016}).

Estimates of stellar mass, however, critically depend on quantities like the initial mass function (IMF), the dust content, the metallicity and the star-formation history (SFH) of each galaxy (see e.g. \citealt{marchesini2009,conroy2009,behroozi2010,dunlop2012}). Indeed, different sets of SED models characterised by e.g., a different treatment of the TP-AGB phase, have been shown to potentially introduce systematics in stellar mass measurements as large as few decimal dex  \citep{muzzin2009, marchesini2010, pforr2012, mitchell2013, grazian2015}.  Similarly, different dust laws (\citealt{cardelli1989,calzetti2000,charlot2000}) and star-formation histories (SFHs) can increase the systematics by up to $\sim0.2$~dex (\citealt{perez-gonzalez2008}). Furthermore, recently, emission from nebular lines has been found to potentially bias stellar mass measurements (\citealt{schaerer2009, stark2009}).  Stellar masses of high redshift galaxies ($z\gtrsim4$) are particularly sensitive to the contamination by nebular lines, since high-z star-forming galaxies are likely to be characterised by emission lines with equivalent width in excess of $\sim$1000\AA~(see e.g., \citealt{labbe2013,debarros2014,smit2014}). Nonetheless, attempts to correct for this contamination can lead to results differing by factors of a few (e.g., \citealt{stark2009,labbe2010b,gonzalez2012,stefanon2015}).

Alternatively, one could directly study the rest-frame optical/NIR light emitted by the low-mass stars in galaxies, given its connection to stellar mass. Measurements of the optical/NIR luminosity have at least three major advantages over measurements of the stellar mass: 1) they are much less sensitive to assumptions about dust modeling; 2) estimates of luminosity are robust quantities, since luminosity can be recovered directly from the observed flux, with marginal-to-null dependence on the best-fitting SED templates, and hence on e.g., SFH or the IMF; and 3) a careful choice of the rest-frame band reduces the contamination by nebular emission, minimizing the requirement of corrections to the fluxes (for instance, nebular emission could contribute up to 50\% of the rest-frame $R$-band luminosity for galaxies at $z\sim8$ - \citealt{wilkins2016}).

The wavelength range spanned by the rest-frame $H$ and $K_\mathrm{s}$ bands is most sensitive to the lower mass stars. Light in bands redder than these is potentially contaminated by emission from the dust torus of AGNs, while light in bluer bands retains information about the recent SFH. The availability of data up to $\lambda_{\mathrm{obs}}\sim8\mu$m from \emph{Spitzer}/IRAC has enabled the study of the evolution of the LF in rest-frame NIR bands ($J, H, K_\mathrm{s}$) up to  $z\sim4$ (e.g., \citealt{cirasuolo2010, stefanon2013, mortlock2016}). 

At  higher redshifts, the choice of the rest-frame band which more closely correlates with the stellar mass must be a trade-off between including contamination from the recent star-formation and performing the luminosity measurements on observational data rather than relying on the extrapolation of SED templates.

In this context, the $z'$ band ($\lambda_\mathrm{eff}\sim0.9\mu$m) emerges as a natural choice: it lays in the wavelength regime redder than the Balmer/4000\AA~break, it is free from contamination by strong nebular emission, and it can be probed up to $z\sim8$ thanks to the \emph{Spitzer}/IRAC $8.0\mu$m-band data.

Recently, \citet{labbe2015} have assembled the first full-depth IRAC mosaics over the GOODS and UDF fields, combining IRAC observations from the  IGOODS (PI: Oesch) and IUDF (PI: Labb\'e) programs with all the available archival data over the two fields (GOODS, ERS, S-CANDELS, SEDS and UDF2).  Following the same procedure implemented by \citet{labbe2015}, full-depth IRAC mosaics have now been generated also for the GOODS-N and COSMOS/UltraVISTA fields. The GOODS-N mosaics double the area with the deepest IRAC imaging available, while the COSMOS/UltraVISTA data, shallower but covering a much larger field, are necessary to  include the most massive galaxies at $z\sim4$. The unique depth of IRAC $5.8\mu$m and $8.0\mu$m mosaics in the GOODS fields (PID 194; PI Dickinson) - reaching $\sim24.5$~AB ($5\sigma$, 2\farcs0 aperture diameter) allowed us to recover flux  measurements with S/N$\gtrsim4$ in these two bands for galaxies to $z\sim7-8$. 

In this work we leverage these characteristics to measure the evolution of the LF at $4 \lesssim z \lesssim 7$ in the rest-frame $z'$ band, providing a complementary approach to the determination of the evolution of the SMF at high redshift. Furthermore, we will show that the SMF can be recovered by applying a stellar mass-to-light ratio ($M_*/L$) to the $z'$-band LF. Remarkably, a simple abundance matching reveals that the $z'$-band LF can also trace the halo mass function (HMF) and its evolution over $4 \lesssim z \lesssim 7$.

Our analysis is based on the photometric catalog of \citet{bouwens2015} over the GOODS-N and GOODS-S fields. At $z\sim4$ we complement our sample with a 37-bands $0.135$-to-$8.0~\mu$m photometric catalog based on the second release (DR2) of the UltraVISTA Survey.  The area covered by the DR2 data, $\sim0.75$ degrees$^2$, is a factor of $\sim10\times$ larger than the cumulative area from the two GOODS fields ($\sim260$arcmin$^2$), enabling the recovery of the bright end of the $z\sim4$ LF with higher statistical significance.  

This paper is organised as follows: in Sect. 2 we describe the adopted datasets and sample selection criteria;  in Sect. 3 we present our results. Specifically, Sect. 3.2 presents the stellar mass-to-light ratios from stacking, while our LF and SMF measurements are presented in Sect. 3.3 and 3.4, respectively. We discuss our LFs measurements with respect to the halo mass function in Sect. 4, and conclude in Sect. 5. Throughout this work we adopted a cosmological model with $H_0=70$ Km/s/Mpc, $\Omega_\mathrm{m}=0.3$, $\Omega_\Lambda=0.7$. All magnitudes refer to the AB system. We assumed a \citet{chabrier2003}  IMF unless otherwise noted.

\section{The sample of $4<\MakeLowercase{z}<7$ galaxies}

\subsection{Data sets}
\label{sect:data_sets}

\begin{figure*}
\includegraphics[width=16cm]{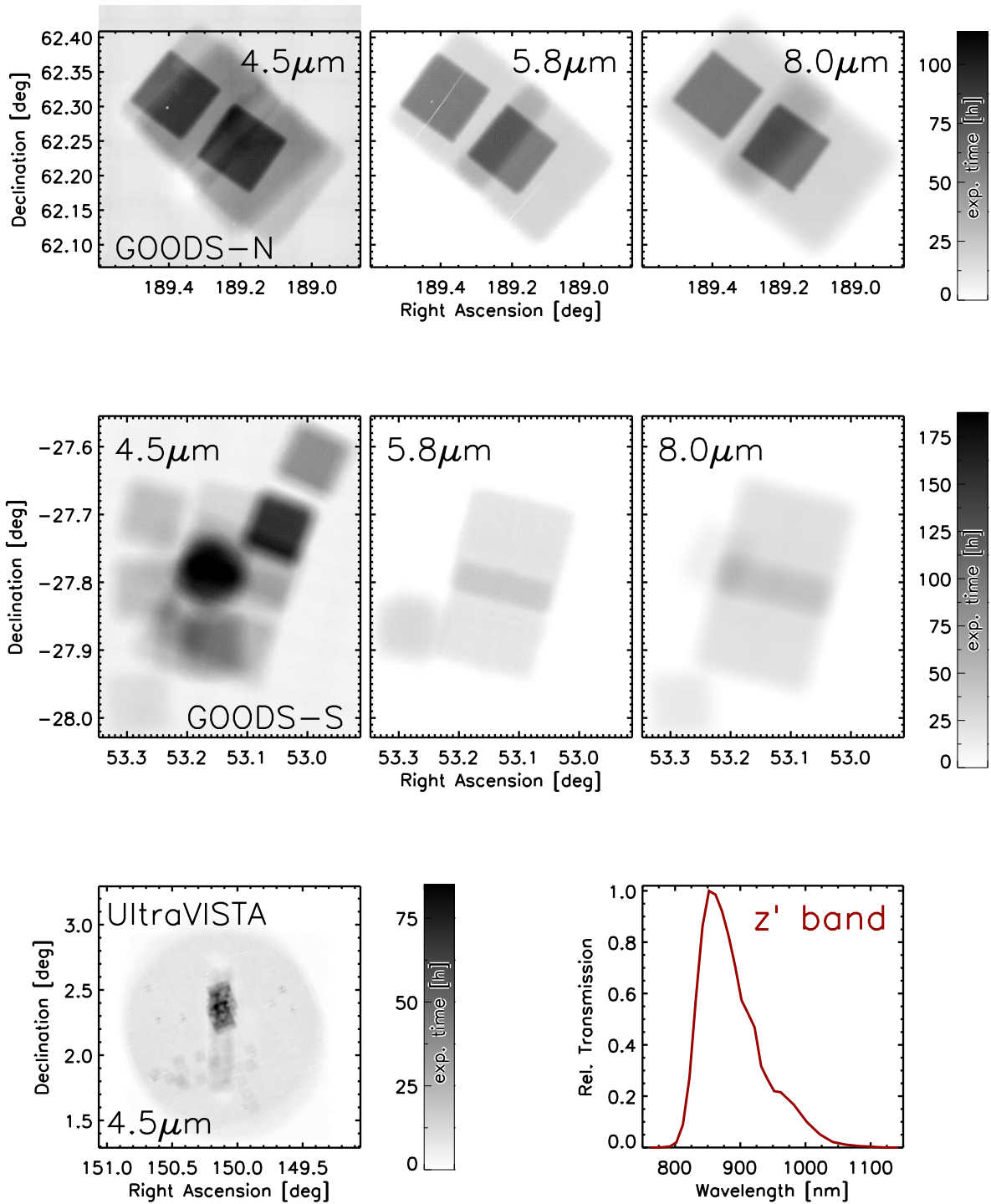}
\caption{Exposure maps of the full-depth IRAC mosaics used in this work for the measurements of the $z'$ band absolute magnitude. The maps are shown with inverted grey scale, and maintain the same scaling stretch across all panels in each row to highlight the relative exposure times; the amount of exposure time is indicated by the vertical bar on the right. Each row refers to a different field: GOODS-N, GOODS-S/ERS and UltraVISTA, top to bottom, respectively, labeled in the left-most panel. Left to right, the panels present the $4.5\mu$m, $5.8\mu$m and $8.0\mu$m mosaics. For the UltraVISTA field, we only show the $4.5\mu$m mosaic, as we use this dataset only at $z\sim4$.  The lower-right panel presents the filter transmission curve for the $z'$ band adopted in this work.\label{fig:exptime_maps}}
\end{figure*}

Our LF measurements are based on a composite sample of galaxies at $4 \lesssim z \lesssim 7$ selected in the rest-frame optical ($z'$ band, $\lambda_\mathrm{eff}\sim0.9\mu$m - see Figure \ref{fig:exptime_maps} for its transmission efficiency), with the bulk of our sample formed by Lyman break galaxies (LBGs) from the CANDELS/GOODS-N, CANDELS/GOODS-S $-$ ERS fields. The $z\sim4$ bin is complemented by a sample of galaxies from a catalog based on the UltraVISTA DR2 mosaics. 

The LBG samples in the CANDELS GOODS-N/S and ERS fields rely on the multi-wavelength photometric catalogs of \citet{bouwens2015}. These are based on the re-reduction of public HST imaging and are enhanced by proprietary full depth \emph{Spitzer}/IRAC mosaics. Specifically, they benefit from novel full depth IRAC $5.8\mu$m and $8.0\mu$m mosaics, not available in the original catalog of \citet{bouwens2015}. The UltraVISTA DR2 catalog is based on the most recent publicly available mosaics at UV-to-NIR wavelengths (including UltraVISTA DR2 data sets) and complemented by an internal release of full-depth \emph{Spitzer}/IRAC mosaics.

In the following sections we briefly describe these two parent catalogs and detail the criteria we adopted to assemble our final sample of galaxies.

\subsubsection{GOODS-N/S and ERS}
 
For this work we adopted the catalog assembled by \citet{bouwens2015} over the CANDELS/GOODS-N, CANDELS/GOODS-S and ERS fields. Here we briefly summarise the main features referring the reader to Sect. 2 and 3 of \citet{bouwens2015}  for full details. 

The catalog contains the photometry in the \emph{HST} ACS F435W, F606W, F775W and F850LP bands (hereafter indicated by $B_{435}, V_{606}, i_{775}$ and $z_{850}$), together with \emph{HST} WFC3 F105W, F125W, F160W ($Y_{105}, J_{125}, H_{160}$) data from CANDELS \citep{grogin2011} and  WFC3 F140W band ($JH_{140}$) from the 3D-HST \citep{brammer2012, skelton2014} and AGHAST (\citealt{weiner2014} - \url{http://mingus.as.arizona.edu/~bjw/aghast/}). 

The catalog takes also advantage of full-depth mosaics in the four \emph{Spitzer}  IRAC bands. The $3.6\mu$m and $4.5\mu$m mosaics were assembled combining data from the IGOODS (PI: Oesch) and IUDF (PI: Labb\'e) programs to all the public archival data from either cryogenic or post-cryogenic programs over the GOODS-N and GOODS-S (GOODS, ERS, S-CANDELS, SEDS and UDF2). For the $5.8\mu$m and $8.0\mu$m mosaics, instead, only observations from the GOODS cryogenic program are available (PI: Dickinson, PID: 194). The mosaics were regenerated from the AORs using the same procedure of \citet{labbe2015}. This procedure delivers the most accurate reconstruction of the PSF at any position across each mosaic, enabling a more accurate flux measurement in the IRAC bands (see below). The mosaics in the $4.5\mu$m, $5.8\mu$m and $8.0\mu$m bands are key for this work as they probe the rest-frame $z'$ band. Specifically, the $4.5\mu$m band matches the rest-frame $z'$ band at $z\sim4$, while the $5.8\mu$m and $8.0\mu$m bands are required for the rest-frame $z'$ band at $z\gtrsim5$.

Figure \ref{fig:exptime_maps} presents the exposure time maps in the IRAC $4.5\mu$m, $5.8\mu$m and $8.0\mu$m for the GOODS-N and GOODS-S fields. As a result of the combination of data from different programs the achieved depth across each field is highly inhomogeneous. This is particularly evident for the $4.5\mu$m band whose depth ranges between 50~hr and 180~hr (corresponding to $25.1-25.8$~AB, respectively, for $5\sigma$, 2\farcs0 aperture diameter).  The GOODS-N field is characterised by the deepest $5.8\mu$m and $8.0\mu$m data, reaching a depth of $\sim80$ hr ($\sim 24.5$~AB, $5\sigma$, $2\farcs0$ aperture).

The object detection was performed on the $\chi^2$ image \citep{szalay1999} constructed from the $Y_{105}, J_{125}, H_{160}$ band images. The detection mosaics have footprints  of $\sim124$ and $\sim140$  arcmin$^2$, respectively for GOODS-N and GOODS-S, for a total of 264 arcmin$^2$. Aperture photometry in the \emph{HST} bands was performed in dual mode with SExtractor \citep{bertin1996} on the mosaics matching the resolution of the $H_{160}$ image. Fluxes were converted to total through the application of an aperture correction based on the \citet{kron1980} scalable apertures and further corrected to take into account the flux losses of the scalable apertures compared to the point-spread function (PSF). Photometry of the IRAC mosaics was performed using a proprietary deblending code (\citealt{labbe2006,labbe2010a,labbe2010b,labbe2013}). This code convolves the high-resolution \emph{HST} mosaics with a kernel obtained from the highest signal-to-noise (S/N) IRAC PSFs to construct a model of the IRAC image. For each object, $2\farcs0$-diameter aperture photometry is performed on the image, previously cleaned from neighbours using the information from the model image. The aperture fluxes were then corrected to total using the HST template specific of each source convolved to match the \emph{Spitzer} IRAC PSF.

Candidate Lyman-break galaxies (LBGs) at $z\sim4,5,6$ and $7$ were selected among the $B_\mathrm{435}, V_\mathrm{606}, i_\mathrm{775}$ and $z_\mathrm{850}$ dropouts, respectively. For a complete list of criteria adopted to select each sample see Table 2 of \citet{bouwens2015}. The sample included 8031 LBGs.

\subsubsection{UltraVISTA DR2}

For the sample of $z\sim4$ galaxies, we also considered detections in the COSMOS/UltraVISTA field, whose larger field compared to GOODS-N/S allowed us to probe higher luminosities.

The UltraVISTA catalog used for this work is based on the \emph{ultradeep} stripes of the second data release (DR2) of the UltraVISTA survey \citep{mccracken2012}. This release is characterised by $5\sigma$ depth  of $\sim 25.6$, 25.1, 24.8, 24.8~AB ($2\farcs0$ aperture diameter) in $Y, J, H$ and $K_\mathrm{s}$, respectively ($\sim0.8-1.2$~mag deeper than DR1) and extends over an area of $\sim 0.75$ square degrees in 4 stripes over the COSMOS field \citep{scoville2007}. The 37-bands catalog was constructed following the same procedure presented in \citet{muzzin2013a} for the DR1.  Briefly, the detection was performed in the \Ks band; 33-band far UV-to-\Ks aperture fluxes were measured with SExtractor (\citealt{bertin1996}) in  dual mode  on the  mosaics matching the PSF resolution of the $H$-band image. An aperture correction recovered from the Kron ellipsoid was applied on a per-object basis; total fluxes were finally computed by applying a further aperture correction obtained from the PSF curve of growth. This new catalog also includes  flux measurements in the  Subaru narrow bands NB711, NB816, the UltraVISTA narrow band NB118 and the CFHTLS $u^*$, $g'$, $r'$, $i'$ and $z'$, not available in the DR1 catalog of \citet{muzzin2013a}.

The COSMOS field benefits from several hundreds hours of integration time with \textit{Spitzer} IRAC. Similarly to what was done for the GOODS-N/S fields, full-depth mosaics were constructed following the procedure of \citet{labbe2015}. Specifically full depth  3.6$\mu$m and $4.5\mu$m mosaics were reconstructed combining data from the S-COSMOS \citep{sanders2007}, S-CANDELS \citep{ashby2015} and SPLASH (PI: Capak, \citealt{steinhardt2014}). The resulting coverage map for the $4.5\mu$m band is shown in the lower panel of Figure \ref{fig:exptime_maps}. The depth ranges from $\sim 4$ to $\sim90$~hours which correspond to $\sim 23.8-25.4$~AB  ($5\sigma$ in a $2\farcs0$ aperture).

Observations in the $5.8\mu$m and $8.0\mu$m channels are only available from the S-COSMOS \textit{Spitzer} cryogenic program. These data have a much shallower depth compared to the $3.6\mu$m and $4.5\mu$m bands, with an average limit of $\sim$22.2~AB ($5\sigma$, 2\farcs0 aperture). For this reason we only considered galaxies from the GOODS-N/S fields for the $z\ge5$ samples.

Fluxes in the four IRAC bands were measured using the template fitting procedure of \citet{labbe2006,labbe2010a,labbe2010b,labbe2013}, adopting the \Ks band as the high-resolution template image to deblend the IRAC photometry.

Photometric redshifts  were computed using EAzY \citep{brammer2008} on the 37 bands photometric catalog, complementing the standard EAzY template set with a maximally red template SED, i.e., an old (1.5~Gyr) and dusty ($A_\mathrm{V}=2.5$~mag) SED template.  We only considered objects whose fluxes were not contaminated by bright nearby stars, had extended morphology on the $K_\mathrm{s}$ image, and less than five bands were excluded as the associated flux measurements were contaminated by \texttt{NaN} values. Galaxies in the $z\sim4$ redshift bin were selected among those with photometric redshift $3.5<z_\mathrm{phot}<4.5$. The initial $z\sim4$ sample included 1208 objects. 

The photometric redshift selection allowed us to consider objects which could have been missed by a pure LBG selection. The large area offered by the UltraVISTA DR2 footprint enabled the selection of bright/luminous sources whose surface density would be too low to be probed over the GOODS-N/S fields area. Such luminous systems could be intrinsically redder than normal LBGs either because they are more (massive) evolved systems and/or they contain a higher fraction of dust.  On the other side, the LBG selection at fainter luminosities from the GOODS-N/S samples is expected to suffer only limited selection bias against intrinsic red sources as in this range of luminosities galaxies are mostly blue star-forming systems, with low content of dust.

\begin{figure}
\hspace{-1cm}\includegraphics[width=10cm]{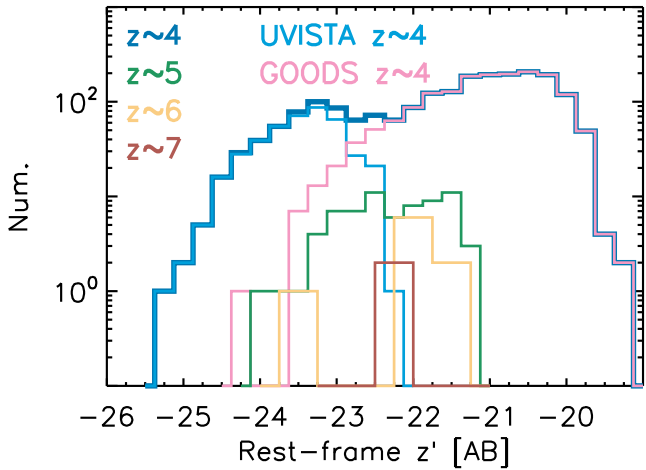}
\caption{Distribution of the absolute magnitudes in the $z'$ band of our composite sample of galaxies at $z\sim4, 5, 6$ and $7$, as labeled in the figure. For the $z\sim4$ sample,   the histograms for the GOODS-N/S and UltraVISTA samples are also presented separately,  showing the complementarity in $M_{z'}$ of the two datasets. \label{fig:absM_histogram}}
\end{figure}

\subsection{Sample assembly}
\label{sect:sample_assembly}

The first step consisted in applying a cut in the $K_\mathrm{s}$ flux of the galaxies from the UltraVISTA sample, in order to control the detection completeness. The DR2 data is $\sim1$~mag deeper than DR1. Therefore, we set the threshold to $K_\mathrm{s}=24.4$~mag, corresponding to the 90\% completeness in detection of point-sources \citep{muzzin2013a}. Instead, we did not apply any cut in the detection band of the GOODS-N/S sample, as the method adopted to estimate the co-moving volumes already takes into account the incompleteness from the detection stage. 

Successively, we excluded from our sample those galaxies with poor flux measurements in those IRAC bands used to compute the rest-frame $z'$ luminosity. The variation in depth across each IRAC mosaic prevented us from applying a single value of flux threshold at this stage. Instead, we applied a cut in S/N to the flux in the IRAC band closest to the rest-frame $z'$ band (i.e.,  IRAC $4.5\mu$m at $z\sim4$ and IRAC $5.8\mu$m and $8.0\mu$m at $z\gtrsim5$). 

Considering the gap in the photometric depth probed by the $4.5\mu$m mosaics compared to that reached by the $5.8-8.0\mu$m data, we opted for applying a distinct S/N cut depending on the considered redshift bin.  The sample at $z\sim4$ was selected by applying the cut of $S/N>5$ to the $4.5\mu$m flux; the samples at $z\sim5,6,7$ were assembled by considering the cumulative flux in the $5.8\mu$m and $8.0\mu$m bands as the inverse-variance weighted sum of the flux in these two bands. We then applied a cut to the corresponding S/N such that:
\begin{equation}
\mathrm{S/N}_{5.8+8.0}\equiv \frac{\mathrm{S}_{5.8} w_{5.8}+{\mathrm{S}_{8.0} w_{8.0}}}{\sqrt{w_{5.8}+w_{8.0}}}>4
\end{equation}
where S$_\kappa$ is the flux measurement in band $\kappa$ and $w_\kappa$ is the weight defined as $1/\sigma^2_\kappa$, with $\sigma_\kappa$ the corresponding  flux uncertainty. The application of the S/N cut reduced the number of galaxies to  2644 (2040/604 for GOODS/UltraVISTA, respectively), 96, 17 and 4 at $z\sim4$, $z\sim5$, $z\sim6$ and $z\sim7$, respectively.

We further cleaned our sample, excluding those objects satisfying any of the following conditions: 1) the contribution to the $5.8\mu$m and $8.0\mu$m flux from neighbouring objects is excessively high; 2) the source morphology is very uncertain or confused making IRAC photometry undetermined; 3) the source is detected at X-rays wavelengths, suggesting it is a lower redshift AGN; 4) the source is at higher redshift, but its SED is dominated by AGN light; 5) LBGs with a likely $z<3.5$ solution from photometric redshift analysis. In Appendix \ref{app:sel}, we detail our application of these additional criteria in cleaning our sample. 

Our final sample consists of 2098 galaxies at $z\sim4$ (1680 from the LBG sample and 418 from the UltraVISTA sample), 72 at $z\sim5$, 10 at $z\sim6$ and 2 objects at $z\sim7$. The distribution of the absolute magnitudes in the $z'$ band for the sample is presented in Figure \ref{fig:absM_histogram}, for the four different redshift bins. It is noteworthy how the GOODS-N/S and UltraVISTA samples complement each other at $z\sim4$, allowing to fully exploit these data with little redundancy. 

\subsection{Selection biases}
\label{sect:biases}

The samples adopted in this work rely on LBG selection criteria, complemented at $z\sim4$ by a photometric redshift selected sample based on the UltraVISTA DR2 catalog.  

The criteria adopted for the assembly of our samples introduce two potential biases to our estimates of LF, $M_*/L$ ratio and SMF (\citealt{fontana2006, grazian2015}). The Lyman Break criteria select, by construction, blue star-forming galaxies, and may thus exclude a greater fraction of red objects compared to photometric-redshift selections. Furthermore, even samples based on photometric redshifts can suffer incompleteness from very red sources, too faint to appear in the detection bands (usually $H$ or $K_\mathrm{s}$ or a combination of NIR bands), but that emerge at redder wavelengths (e.g., IRAC). In the following we attempt to evaluate the impact of these biases on our sample.

\begin{figure}
\hspace{-1cm}\includegraphics[width=10cm]{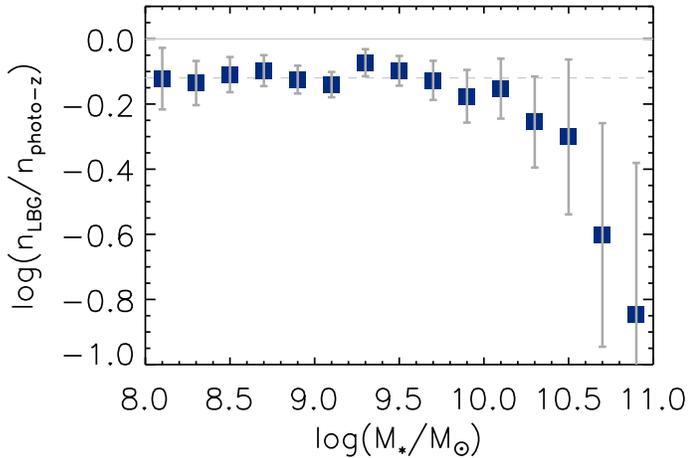}
\caption{Fraction of $z\sim4$ galaxies recovered using LBG criteria relative to the underlying sample of galaxies selected to have photometric redshifts $3.5<z_\mathrm{phot}<4.5$, shown as a function of stellar mass. The horizontal dashed line marks the mean of the recovered fraction for $\log(M_*/M_\odot)\sim10.2$. The LBG criteria recover $\gtrsim75\%$ of all the sources up to $\log(M_*/M_\odot)\le10.2$. \label{fig:lbg_photoz}}
\end{figure}

From the stellar mass catalog of CANDELS/GOODS-S (\citealt{santini2015}), we extracted those galaxies with photometric redshift $3.5<z_\mathrm{phot}<4.5$. Successively, we applied the LBG criteria to their flux measurements from the multi-wavelength catalog of \citet{guo2013}.  In Figure \ref{fig:lbg_photoz} we present, as a function of stellar mass, the ratio between the number of galaxies recovered through the LBG criteria and the number of galaxies in the photo-z sample. The plot shows that the LBG  selection is able to recover $\gtrsim75\%$ of galaxies  with stellar mass $\log(M_*/M_\odot)\lesssim10$ (corresponding to $\sim-0.12$~dex offset in number density measurements); at $\log(M_*/M_\odot)\lesssim10.5$ the galaxies missed by the LBG criteria amount to about 35\% ($\sim0.2$~dex). At higher stellar masses, the fraction of galaxies not entering the Lyman Break selection increases to $\gtrsim60-70\%$ ($\sim0.5-0.6$~dex) consistent with \citet{grazian2015}. \citet{duncan2014} showed that photometric uncertainties scatter a large fraction of the measurements outside the LBG selection box; specifically, the LBG criteria recover only $\sim1/4$ (equivalent to $\sim-0.6$~dex offset) of the galaxies recovered through photo-z (see also \citealt{dahlen2010}). However, once selection criteria on the redshift probability distribution are introduced, excluding from the sample poorly constrained photometric redshifts, the resulting photo-z sample largely overlaps with the LBG one, as demonstrated by the fact that the resulting photometric redshift UV LFs agree well, usually within $1-\sigma$, with the LBG UV LF (\citealt{duncan2014, finkelstein2015a}).

The photometric depth of the UltraVISTA DR2 catalog, $K_\mathrm{s}=24.4$~mag, corresponds to a stellar mass completeness limit for a passively evolving simple stellar population of $\log(M_*/M_\odot)\sim10.6$ at $z\sim4$ and $11.2$ at $z\sim5$. The depth in the GOODS-Deep fields correspond to limits in stellar mass of $\log(M_*/M_\odot)\sim10.3$ at $z\sim4$ and $10.6$ at $z\sim5$, respectively. We would like to remark that our analysis for the $z\sim4$ sample at stellar masses $\log(M_*/M_\odot)\gtrsim10-10.5$ is dominated by the photometric redshift sample from UltraVISTA, covering a larger volume for bright sources than the GOODS fields. Therefore, our composite $z\sim4$ sample is only marginally affected by the LBG selection bias.

A number of works have studied the so called extremely red objects, characterized by very red ($\gtrsim2-3$~mag) rest-UV/optical colors, making them more elusive in high-z samples (e.g., \citealt{yan2004,huang2011,caputi2012, stefanon2015, caputi2015, wang2016}). Samples detected in IRAC bands suggest that many of these objects could be consistent with being $z\gtrsim3$ massive galaxies.

Recently, \citet{wang2016} analyzed the properties of $H-[4.5]>2.25$~mag over the CANDELS/GOODS-N and GOODS-S fields. Interestingly they identified 18 sources not present in the $H_\mathrm{160}$-band catalog, but included in the IRAC catalog of \citet{ashby2013}. Of these, 5 sources have an estimated photometric redshift $3.5<z_\mathrm{phot}<4.5$ and have a stellar mass $10.5 \lesssim \log(M_*/M_\odot) \lesssim 11$. Since their analysis refers to the same fields we consider in our work (although likely the  configurations at the detection stage are different), we can use their result to obtain a rough estimate of the fraction of objects missed by our selection. Assuming a $M_*/L_{z'}\sim0.2 M_\odot/L_\odot$, quite typical for these masses and redshifts (as we show in Section \ref{sect:ml_ratios}), the stellar mass range of these galaxies would correspond to luminosities  $-24.7\lesssim M_{z'} \lesssim -23.4$~AB. This sample would then constitute $\sim65\%$ of the objects in our $z\sim4$ LBG sample with similar luminosities. This fraction drops to $\sim8\%$ when comparing the 5 sources to the $\sim60$ galaxies with the same photometric redshift and stellar mass over the CANDELS/GOODS fields. 

 \citet{caputi2015} presented SMF measurements at $z\sim3-5$ obtained complementing the SMF from a photometric-redshift, $K_\mathrm{s}$-selected sample based on UltraVISTA DR2 data to SMF measurements from photometric redshift samples of $K_\mathrm{s}$-dropouts detected in IRAC bands. The main result is that $K_\mathrm{s}$ dropouts can account for as high as $\sim0.5$~dex in number densities. However, \citet{stefanon2015} showed that samples similar to those of \citet{caputi2015} likely suffer from degeneracies in the measurement of photometric redshifts (and consequently stellar masses), and therefore the above estimate is still uncertain.

The depth of current IRAC data, however, is not sufficient to systematically inspect passively evolving stellar population with stellar masses below $\sim10^{10-10.5} M_\odot$ at $z\gtrsim4$. We therefore caution the reader that any  sample currently available dealing with stellar mass below the $\sim10^{10} M_\odot$ limit may still be biased against dusty and/or old galaxies. Forthcoming projects, like \emph{Spitzer}/GREATS (Labb\'e et al., in prep.) and the JWST will allow us to obtain a more complete picture by probing the lower mass regime.

\subsection{Selection efficiency and completeness}
\label{sect:completeness}

\begin{figure}
\includegraphics[width=9cm]{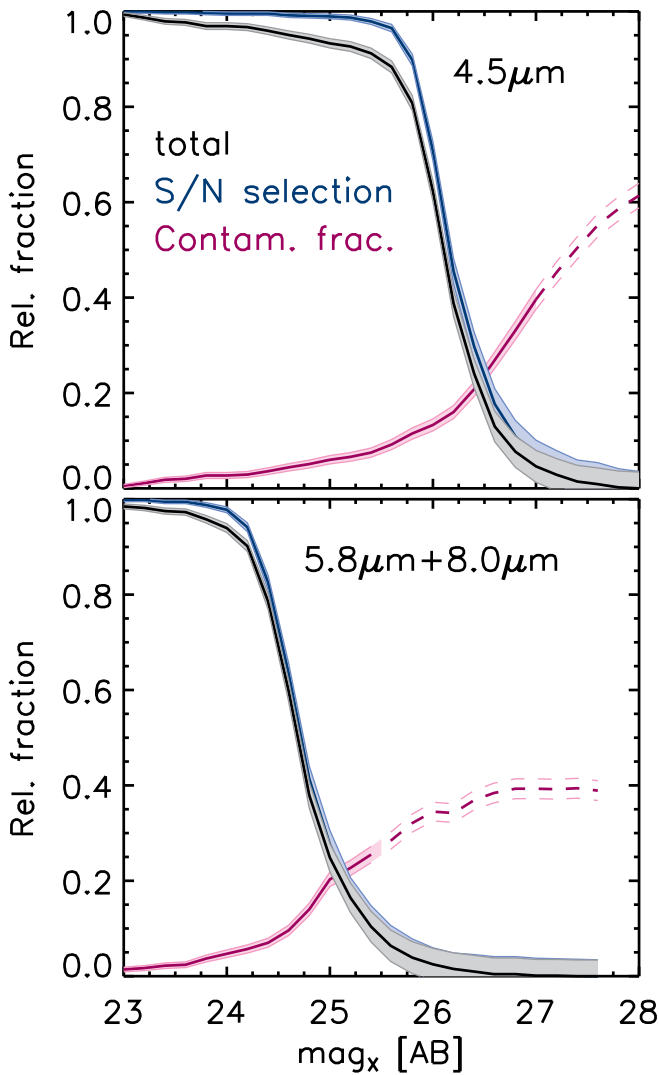}
\caption{Completeness fraction as a function of apparent magnitude from our Monte Carlo simulation for the selection in the $4.5\mu$m (top panel) and $5.8\mu$m+$8.0\mu$m (bottom panel) bands. The pink curve presents the fraction of objects excluded because their flux measurements were highly contaminated by neighbours. The pink shaded area presents the associated $1 \sigma$ Poisson uncertainties. The blue curve and shaded area present the completeness fraction from the S/N cut in the flux of the corresponding band and associated Poisson uncertainty, while the black curve and grey shaded area show the combined effect of S/N threshold and contamination cleaning. \label{fig:completeness}}
\end{figure}

We implemented a Monte Carlo simulation based on real data to estimate the effects that our selection criteria in S/N and contamination polishing have on the sample of galaxies used in this work.  For this simulation, we did not consider the effects of selection in the detection band because the UltraVISTA sample is 90\% or more complete in $K_\mathrm{s}$ by construction, while the effects of detection completeness of the GOODS-N/S sample have been taken into account when estimating the co-moving volumes adopted for the LF measurements.

At first we defined a grid in apparent magnitude of width 0.20~mag. Given the small sizes of the galaxies compared to the IRAC PSF, for each magnitude value in the grid, we injected 100 point sources randomly distributed across a region of uniform depth in the $4.5\mu$m, $5.8\mu$m and $8.0\mu$m mosaics of the GOODS-N field. We chose the GOODS-N as this field is characterised by the deeper \textit{Spitzer}/IRAC $4.5\mu$m, $5.8\mu$m and $8.0\mu$m band data among the fields considered for this work. Successively, we replicated the flux measurement using the same procedure adopted for the actual photometry. 

The completeness fraction in each magnitude bin was computed by comparing the number of objects satisfying our selection criteria (Section \ref{sect:sample_assembly}) to the number of objects initially injected into the simulation.  For the completeness of the $z\sim4$ sample, the above process was applied to the $4.5\mu$m mosaic only.  For the completeness of the samples at $z\ge5$, the point sources were added at matching positions in the $5.8\mu$m and $8.0\mu$m mosaics. The selection on the S/N and contamination was then recovered applying the corresponding criteria and assuming the SED to be flat in $f_\nu$  in the observed $5-9\mu$m region. This is a reasonable approximation since, as we show in Section \ref{sect:ml_ratios}, the median SEDs do not substantially deviate from a flat  $f_\nu$ SED in the wavelength range covered by IRAC observations. The whole process was repeated $10\times$ in each band in order to increase its statistical significance. The global completeness (i.e., the cumulative effects of S/N and contamination selection) at the different depths of the IRAC mosaics was obtained rescaling the completeness in S/N selection to match the depth of the relevant region.

The results from our completeness simulation for the $4.5\mu$m and the $5.8\mu$m+$8.0\mu$m samples are presented in Figure \ref{fig:completeness}. First we discuss the recovery of the contamination fraction; successively we consider these results in the budget of the global completeness estimates.

The fraction of objects in the $4.5\mu$m band contaminated by neighbours\footnote{We consider a flux measurement to be contaminated when 65\% or more of the flux at the position of the source comes from neighbouring objects. See Appendix A for further details.} is negligible for objects brighter than $\sim24$~AB and increases exponentially up to $\sim27$~AB, where it starts to flatten out. A similar behaviour is observed for the $5.8\mu$m+$8.0\mu$m simulation, although shifted at brighter magnitudes, due to the shallower depth of the $5.8\mu$m and $8.0\mu$m compared to the $4.5\mu$m. The flattening at the faint end is caused by a strong incompleteness in the data  at such faint magnitudes and likely does not reflect the true behaviour. In what follows and in our analysis we do not consider the completeness for magnitudes fainter than those corresponding to the onset of the flattening, i.e., $\sim27$~AB and $\sim25$~AB for the $4.5\mu$m and $5.8\mu$m+$8.0\mu$m data respectively.

As it could intuitively be expected, the bright end of the global completeness curve is dominated by the (small) fraction of purged objects. This effect becomes less and less pronounced at fainter magnitudes, corresponding to lower S/N, where the effective selection is driven by the S/N itself.

\begin{figure*}
\includegraphics[width=18cm]{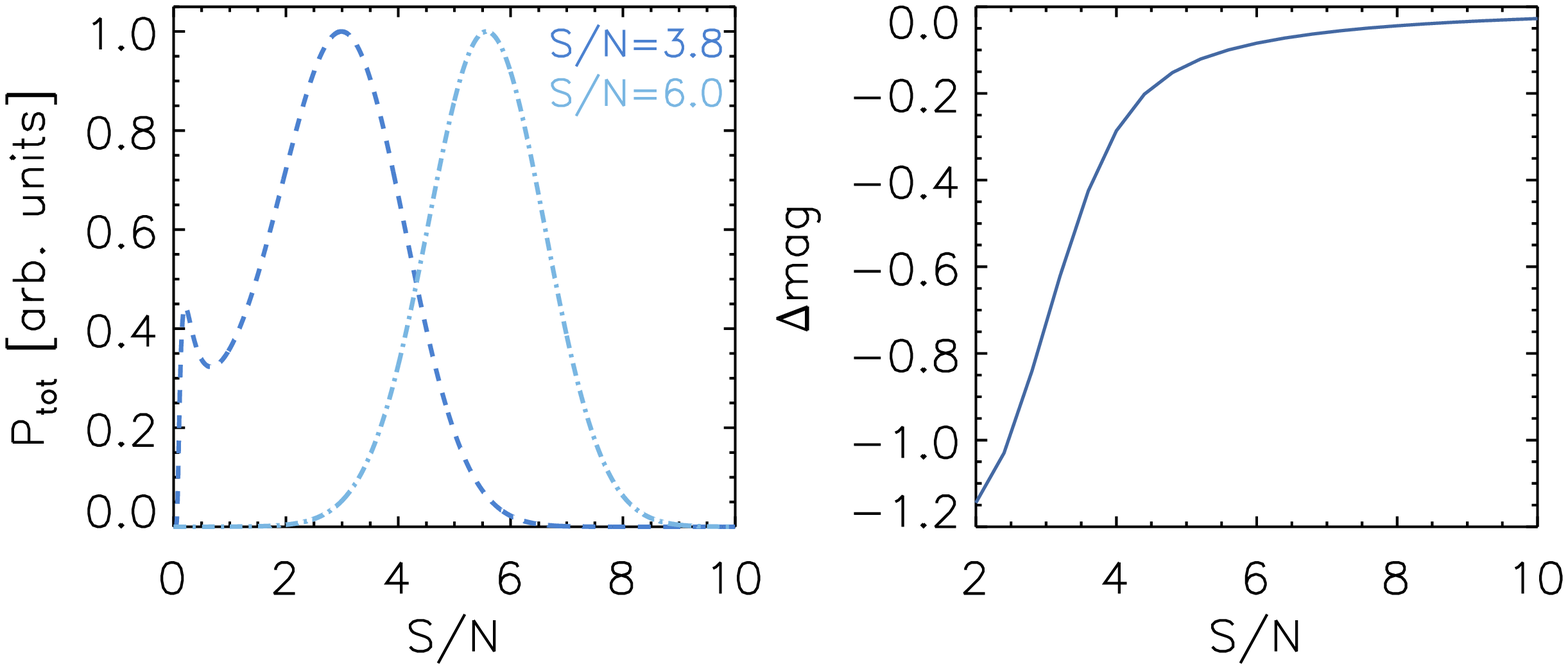}
\caption{{\bf Left panel:} Examples of probability distribution of the intrinsic flux $\mathcal{P}(f_i)$, presented as a function of S/N, for  two cases of \emph{observed} S/N=3.8 and 6.0, as labeled at the top-right. The probabilities have been arbitrarily re-normalized to a maximum of 1, for ease of readibility. {\bf Right panel:} Expected flux boosting as a function of S/N, resulting from Eq. \ref{eq:P(f_i)}. The flux boosting for S/N$\lesssim2.5$ is $\gtrsim1$~mag, suggesting that the recovery of the intrinsic flux for these cases can be very uncertain. \label{fig:flux_boost}}
\end{figure*}

\subsection{Flux boosting}

The random noise from the background can positively combine at the location of a given source, introducing an increase in the measured flux (\emph{flux boosting}, \citealt{eddington1913}). The amount of this boost is inversely correlated to the S/N of the source. The flux for sources with very high S/N will mostly be the result of the photons emitted by the source itself, with reduced contribution from the background; on the other hand, for sources with low S/N, the background level can be typically just few factors smaller than the intrinsic signal from the source, making it sensitive to (positive) fluctuations of the background.  Furthermore, sources do not uniformly distribute with flux, but rather follow a $\sim$ power-law, with fainter sources more numerous than brighter ones. Therefore it is intrinsically more probable that fainter sources scatter to brighter fluxes than the reverse, giving origin to a net flux bias.

A second potential source of flux boosting comes from confusion noise: faint sources at apparent positions close to a brighter one are more likely to be blended into the brighter source, increasing the flux, and decreasing the number of fainter objects. This effect is larger for flux measurement in those bands with wide PSF, like \textit{Spitzer}/IRAC. However, in our case, the photometry in the IRAC bands was performed adopting a higher resolution morphological prior from \emph{HST} mosaics (see Sect. \ref{sect:data_sets}). Furthermore, we applied a selection in flux contamination (see Sect. \ref{sect:sample_assembly}). Since these two factors drastically limit the potential contribution of confusion noise to the IRAC fluxes in our sample, we do not further consider its effects to the flux boosting budget.

For each source, we estimated the flux bias as the ratio between the expected intrinsic flux $f_i$ and the measured flux $f_o$. Since no direct measurement is possible, the intrinsic flux was recovered as the average flux obtained from an estimate of its probability distribution. This was constructed considering two distinct contributions: 1) the probability $\mathcal{P}(f_i, f_o)$ that the observed flux $f_o$ is drawn from the distribution of intrinsic flux $f_i$, given the noise $\sigma_o$, and 2) the frequency $\mathcal{P}(f_i)$ of occurrence of the intrinsic flux $f_i$. Assuming each probability is normalized to 1, the final probability distribution would then be $\mathcal{P}_\mathrm{tot}(f_i)=\mathcal{P}(f_i, f_o)  \mathcal{P}(f_i)$\footnote{A similar expression could also be recovered, modulo a normalization factor, by applying the Bayes's theorem - see e.g. \citet{hogg1998}.}.

Assuming a gaussian noise, $\mathcal{P}(f_i, f_o)$ can be written as:
\begin{equation}
\mathcal{P}(f_i, f_o)=\frac{1}{\sqrt{2\pi\sigma_o^2}}\exp{[-(f_i-f_o)/2\sigma_o^2]},
\label{eq:gaussian}
\end{equation}
normalized to a total probability of 1. 

The frequency associated to the intrinsic flux can be recovered from the (intrinsic) differential number count of sources, $dN(f_i)/df_i$. This can usually be described by a power-law form with negative index, thus divergent for $f_i\rightarrow0$, which prevents it from being normalized\footnote{For S/N $\gtrsim5$, the product between $\mathcal{P}(f_i,f_o)$ and $dN/df$ does not diverge for $f_i\rightarrow0$. However, this is not anymore the case for lower S/N values, where the non-negligible probability of the low-flux tail from the gaussian distribution makes the divergent power-law dominate over the gaussian.}  (see e.g., \citealt{hogg1998}, who also discuss possible reasons for why the divergence at $f_i\rightarrow0$ is likely non-physical).

We therefore followed the formalism of \citet{crawford2010}, who introduced as further constraint the poissonian probability that no sources brighter than $f_i$ exist at the same location of the observed object. The expression for $\mathcal{P}(f_i)$ then becomes:
\begin{equation}
\mathcal{P}(f_i)=\frac{dN(f_i)}{df_i}\times\exp \left( -\Delta\Omega_o\int_{f_i}^{+\infty}dN/df \right)
\label{eq:P(f_i)}
\end{equation}
where $\Delta\Omega_o$ corresponds to the area occupied by the source. In the left panel of Figure \ref{fig:flux_boost} we show examples of reconstructed $\mathcal(P)_\mathrm{tot}$, for the cases of S/N=6 and S/N=3.8, where it is evident the increasing contribution of the faint sources population to the expected intrinsic flux as the S/N decreases. The right panel of Figure \ref{fig:flux_boost} shows the expected flux boost as a function of S/N. For S/N$\gtrsim4.5$, the flux boost is roughly the same amount as the flux uncertainty. However, for lower S/N the estimated flux boost increases abruptly. For S/N$\lesssim2$ the expected flux boost is $\gtrsim1.5$~mag, meaning that the recovery of the intrinsic flux for such low S/N data is highly uncertain. 

The S/N in the \emph{HST} bands for the galaxies in our sample is $>10$. At $z\sim4$ the S/N in the $4.5\mu$m band, adopted for the selection of the $z\sim4$ sample, is $\ge5$ by construction; the S/N in the $3.6\mu$m band is $\ge5$ as well, consistent with the nearly flat SEDs in that wavelength range. At $z\ge5$, the selection was performed in a combination of $5.8\mu$m and $8.0\mu$m fluxes, adopting a S/N$>4$ threshold. Figure \ref{fig:flux_boost} shows that the expected flux boost for S/N$>5$ is $\lesssim0.1$~mag. However, for lower S/N, typical of the selection of samples at $z\ge5$, the correction can be as high as $\sim0.8-1.0$~mag. 

We therefore applied the above correction to the fluxes in the $5.8\mu$m and $8.0\mu$m of the $z\ge5$ samples. The average flux boost was $\sim0.19$~mag and $\sim0.25$~mag in the IRAC $5.8\mu$m and $8.0\mu$m bands, respectively. 

In Appendix \ref{app:SEDs} we present the SEDs of the 12 most luminous galaxies in the $z\sim5$ sample together with the SEDs of the $z\sim6$ and $z\sim7$ samples, before and after applying the flux boost correction.

\section{Results}

\subsection{UV to optical luminosities}

\label{sect:uvopt}

In the last few years a number of works have studied the relation between the rest-frame UV luminosity and the stellar mass of high-redshift galaxies (e.g., \citealt{stark2009,lee2011,gonzalez2011,mclure2011,spitler2014,duncan2014,grazian2015,gonzalez2016,song2016}). Indeed, a relation between the stellar mass and the UV luminosity is to be expected considering a continuous star formation. Deviations from such a relation would then provide information on the age and metallicity of the stellar population and on the dust content of the considered galaxies. 

The emerging picture is that at $z\sim4$ and for stellar masses $\log(M_*/M_\odot)\lesssim 10$, the stellar mass increases monotonically with increasing UV luminosity; however, at stellar masses higher than $\log(M_*/M_\odot)\sim10$ the trend becomes more uncertain: \citet{spitler2014} using a sample of $K_\mathrm{s}$-based photometric redshift selected galaxies found indication of a turnover of the UV luminosity, with the more massive galaxies ($10.5\lesssim \log(M_*/M_\odot)\lesssim 11$) spanning a wide range in UV luminosities (see also \citealt{oesch2013}); \citealt{lee2011}, instead, using a LBG sample, found a linear relation between UV luminosities and stellar masses up to $\log(M_*/M_\odot)\sim 11$. Considering the different criteria adopted by the two teams for the assembly of their samples, selection effects might be the main reason for the observed tension.   

This observed discrepancy could however just be the tip of an iceberg. Indeed current high-z surveys might still be missing lower-mass intrinsically red galaxies (dusty and/or old), that could populate the $M_*-M_\mathrm{UV}$ plane outside the main sequence (\citealt{grazian2015} and our discussion in Sect. \ref{sect:biases}).  The depth of the current NIR surveys does not allow us to further inspect this, which will likely remain an open issue until JWST.

A monotonic relation between the UV luminosity and the stellar mass has also been found at $z\sim5$ and $z\sim6$ for $\log(M_*/M_\odot)\lesssim 10$ (e.g., \citealt{stark2009,gonzalez2011,duncan2014,song2016, salmon2015}), with approximately the same slope and dispersion, but with an evolving normalisation factor (but see e.g.,  Figure 5 of \citealt{song2016} for further hints on the existence of massive galaxies with faint UV luminosities).

\begin{figure}
\includegraphics[width=9cm]{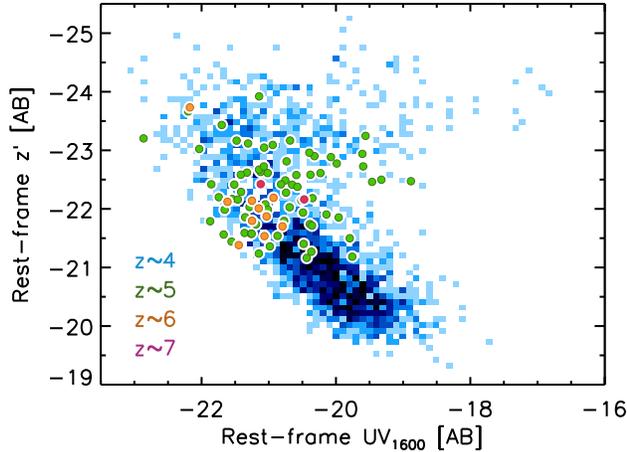}
\caption{$z'$ absolute magnitudes versus the UV absolute magnitudes for our composite sample, color-coded according to the considered redshift bin. The $z\sim4$ data are presented as a density plot, with denser regions identified by a darker color, while the points for the $z\ge5$ samples are shown individually. The UV-$z'$ relation shows a turnover for $M_{z'}\lesssim-22.5$. \label{fig:uvopt}}
\end{figure}

\begin{figure}
\hspace{-1cm}\includegraphics[width=9.5cm]{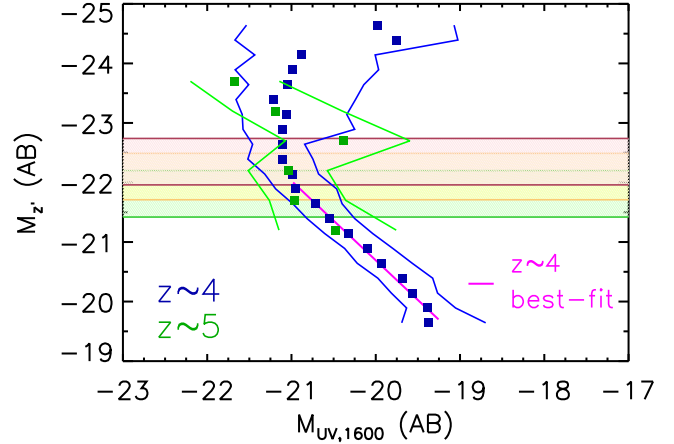}
\caption{Median of the $M_\mathrm{UV}$ vs. $M_{z'}$ relation in bins of $M_{z'}$. The blue points mark the median at $z\sim4$, while the green points mark the median at $z\sim5$. The left and right blue (green) curves represent the 25 and 75 percentiles, respectively, of the points at $z\sim4$ ($z\sim5$). The horizontal green, yellow and red shaded regions indicate the limiting magnitude corresponding to our S/N cuts at $z\sim5$, $z\sim6$ and $z\sim7$, respectively. The magenta line represents the best-fit relation for the $z\sim4$ sample in the range $-22 <M_{z'}<-19.75$~mag. \label{fig:median_uvopt}}
\end{figure}

Figure \ref{fig:uvopt} presents the absolute magnitude in the rest-frame  $z'$ band ($M_{z'}$) as a function of the absolute magnitude in the UV ($M_{\mathrm{UV_{1600}}}$), for our sample in the four redshift bins ($z\sim4,5,6$ and $7$), while in Figure \ref{fig:median_uvopt} we present the binned median in the $M_{\mathrm{UV}}-M_{z'}$ plane for the $z\sim4$ and $z\sim5$ samples. 

The $z\sim4$ sample shows a clear correlation between the luminosities in the rest-frame UV and $z'$ bands for $M_{z'}\gtrsim -22$~mag, which can be described by the following best-fitting linear relation:
\begin{equation}
M_\mathrm{UV}=(-3.58 \pm 1.49) + (0.79 \pm 0.07) \times M_{z'}
\label{eq:uvopt}
\end{equation}
The above best-fit is marked by the magenta line in Figure \ref{fig:median_uvopt}, where we also indicate the $3\sigma$ limits corresponding to our $5.8\mu$m$+8.0\mu$m selection. At $z\sim4$ and  $z\sim5$ the depth of the IRAC data allows us to not only probe the bright end, where the relation between $M_\mathrm{UV}$ and $M_{z'}$ breaks, but to also explore the regime of the linear correlation expressed by Eq. \ref{eq:uvopt}.   Slopes of 0.4-0.5  in the $\log(M_*)-M_\mathrm{UV}$ plane (with nominal $1\sigma$ uncertainties of $\sim0.05-0.1$) have been reported by e.g., \citet{duncan2014} and \citet{song2016}. Assuming a constant $M_*/L_{z'}$ ratio (see Section \ref{sect:ml_ratios}), our measurements correspond to a slope of 0.5 in the $\log(M_*)-M_\mathrm{UV}$ plane, consistent with previous measurements. 

Assuming that the SFR mostly comes from the UV light and that $M_{z'}$ is a good proxy for stellar mass measurements, we can also compare the slope we derived for the $M_\mathrm{UV}-M_{z'}$ relation to that of the log(SFR)-$\log(M_*)$ from the literature.  Indeed, the observed UV slopes of $M_z'\gtrsim -22$ galaxies in our sample are $\beta\sim-2$ (see also \citealt{bouwens2010}), consistent with star forming galaxies and little-to-no dust extinction. Our measurement is perfectly consistent with the log(SFR)-$\log(M_*)$ slope of $\sim0.8\pm0.1$ recently measured by e.g., \citet{whitaker2015} for $z\lesssim 2.5$ star forming galaxies with $\log(M_*/M_\odot)\lesssim 10.5$, and it has been shown to evolve little over the redshift range $0.5 < z < 2.5$.  

For absolute magnitudes brighter than $M_{z'}\sim-22$~mag the linear relation expressed by Eq. \ref{eq:uvopt} breaks, as we observe the beginning of a turnover in the absolute UV-$z'$  magnitude relation. Remarkably, this behaviour is visible also for the $z\sim5$ sample, which is entirely based on LBG selection. This fact has important consequences for e.g., stellar mass function measurements: samples of galaxies selected at fixed rest-frame UV luminosity are potentially characterised by a wide range of stellar mass.

The absolute UV magnitude of galaxies with $M_{z'}\sim-23.5\pm0.8$~mag  spans the full range of values observed for $M_{z'}\lesssim-22.5$~mag. However, the bulk of values aggregates around the $[M_\mathrm{UV},M_{z'}]\sim[-21.4, -23.5]$~mag region and it is characterised by a large dispersion in $M_\mathrm{UV}$ ($\gtrsim3$~mag). This result is qualitatively consistent with what found by \citet{spitler2014}, assuming a correlation between the absolute magnitude $M_{z'}$ and the stellar mass.  Most of the galaxies with $M_{z'}\sim-23.5$ come from the photometric redshift sample selected from the UltraVISTA catalog. As we will present in Sect.~\ref{sect:ml_ratios}, our measurements of the mass-to-light ratios from stacking analysis show that  galaxies with $M_{z'}\lesssim-23.5$ statistically have stellar masses $\log(M_*/M_\odot)\gtrsim10.6$. The above result then underlines the bias that LBG selections may introduce against massive systems.

At $z\sim5$ our data allows us to inspect the relation only for $M_{z'}$ fainter than $-23$ and $M_{\mathrm{UV,1600}}$ fainter than $\sim-22$.  In this range of luminosities, our $z\sim5$ measurements are roughly consistent with the $z\sim4$ measurements in the same range of luminosities.  The measurements for the $z\sim6$ and $z\sim7$ samples are still consistent with the trends observed at $z\sim4$, although the low number of objects does not allow us to derive any statistically significant conclusion.

\subsection{Stellar mass-to-light ratios from stacking analysis}

\label{sect:ml_ratios}

So far determinations of the stellar mass-to-light ratios ($M_*/L$) for galaxies at $z>4$ have involved the $M_*/L_\mathrm{UV}$ ratio. This quantity is fundamental for our understanding of galaxy formation and evolution as it combines information on the recent  (through the UV luminosity) and on the integrated (through the stellar mass) SFH  (e.g \citealt{stark2009}). Nonetheless, the $M_*/L_\mathrm{UV}$  has been used to recover the stellar mass and SMF of high redshift galaxies with alternating success (see e.g., \citealt{gonzalez2011,song2016}). In this section, instead, we explore for the first time the $M_*/L_{z'}$ properties of galaxies at $z\gtrsim4$. The rest-frame $z'$ luminosity is more sensitive to the stellar mass compared to the UV luminosity for two reasons. While the UV light is emitted by massive, short-lived stars and thus traces the SFH in the past few hundred Myr, the luminosity in the rest-frame optical region mostly originates from lower mass, longer living stars, which constitute most of the stellar mass of galaxies. Furthermore, it is less sensitive to the dust extinction, and hence to the uncertainties in its determination, compared to the UV: for a \citet{calzetti2000} extinction curve, an $A_V=1$~mag gives $A_{\lambda1600}\sim2.5$~mag compared to $A_{\lambda9000}\sim0.5$~mag. 

Since here we are interested more on average trends in the $M_*/L_{z'}$ ratios rather studying it for specific galaxies, we performed our analysis using the median stacked SEDs constructed from our composite sample. Due to the different photometric bands in the catalogs, we performed the stacking of sources separately for sources in the GOODS-N/S and UltraVISTA samples. 

The stacked SEDs were constructed as follows (see also \citealt{gonzalez2011}).  At each redshift interval we divided the galaxies  into sub-samples  according to their $M_{z'}$. The different depths reached by the $4.5\mu$m and $5.8\mu$m$+8.0\mu$m samples resulted in different number of subsamples across the redshift bins.  Under the working assumption of limited variation in both redshift and SED shape in each bin of $M_{z'}$,  and for each \emph{HST} band we took the median of the individual flux measurements. Our assumption is also supported by the fact that the SEDs from stacking are generally characterized by a flat $f_\nu$ continuum at both rest-frame UV and optical regimes. Uncertainties on the median were computed from bootstrap techniques, drawing with replacement  the same number of flux measurements as the number of galaxies in the considered absolute magnitude bin. Before median-combining, the fluxes were perturbed according to their associated uncertainty. The process was repeated 1000 times and the standard deviation of the median values was taken as the final uncertainty. For the IRAC bands, median stacking was performed on the mosaic cutouts centered at the position of each source, previously cleaned from  neighbours. Photometry was performed on the median of the images in apertures of $2\farcs5$ diameter. Total fluxes were then recovered through the PSF growth curve. Uncertainties were computed applying to the image cutouts the same bootstrap technique adopted for the median stacking of the fluxes, as described above. In randomly drawing the image cutouts, we preserved the total exposure time.

\begin{figure*}
\includegraphics[width=18cm]{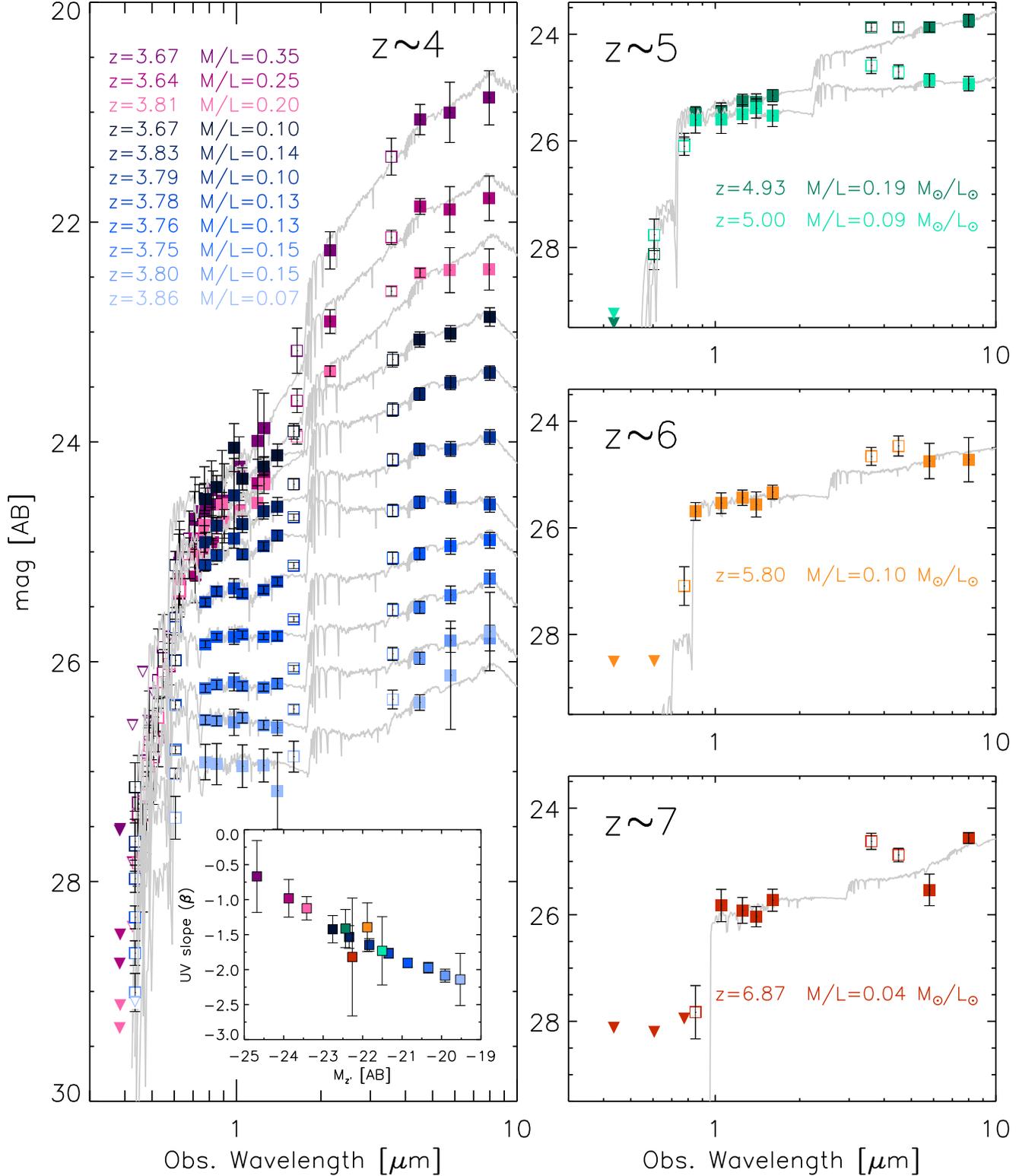}
\caption{Stacked SEDs. Each panel refers to a redshift bin, as indicated by the labels. In each panel, the filled colored squares with error bars represent the SED from the stacking analysis, while the grey curve marks the best-fitting FAST template. The open symbols mark those bands excluded from the fit as potentially contaminated by nebular emission or whose stacked measurement was considered unreliable due to the presence of either the Lyman or Balmer break (see the main text for details). For the $z\sim4$ sample, the stacked SEDs from the UltraVISTA catalog are plotted with shades of magenta, while the stacks from the GOODS-N/S sample are plotted with shades of blue.  In each panel the photometric redshift from EAzY and the mass-to-light ratio ($M/L$) in units of $M_\odot/L_{\odot,z'}$ for each stacked SED are also reported and share the color of the corresponding SED.  The inset in the $z\sim4$ panel presents the relation between the UV slope ($\beta$) and the absolute magnitude $M_{z'}$ for the stacked SEDs in the four redshift bins. Colors match the redshift and luminosity bin.\label{fig:stack}}
\end{figure*}

Photometric redshifts and $z'$-band luminosities were obtained from EAzY (\citealt{brammer2008}) on the stacked SEDs. Briefly, at first EAzY selects the two SED templates that provide the closest match to the observed color in the two filters bracketing the rest-frame band of interest. The luminosity is then computed from the interpolation of the two colors, relative to the rest frame band of interest, obtained from the two selected SEDs (for full details see Appendix C of \citealt{rudnick2003}). Stellar masses were computed running FAST (\citealt{kriek2009}) adopting the \citet{bruzual2003} template SEDs, a \citet{chabrier2003} IMF, solar metallicity and a delayed-exponential SFH.  The bands potentially contaminated by nebular emission were excluded from the fit. Since we performed the stacking in each band individually, assuming the same redshift for all sources, the flux in those bands close to the Lyman and the Balmer breaks potentially suffers from high scatter, introduced by the range of redshifts of the galaxies in each sub-sample, and depending on whether the break enters or not the band. Fluxes in these bands were therefore excluded from the fit with FAST. Specifically, for the $z\sim4$ LBG stacks, we excluded the $B_{435}, V_{606}$ and $H_{160}$ bands; for the $z\sim4$ UltraVISTA stacks, we excluded the $B, \mathrm{IA}427, \mathrm{IA}464, \mathrm{IA}484, \mathrm{IA}505,$ $\mathrm{IA}527, \mathrm{IA}574, \mathrm{IA}624, \mathrm{IA}679,$ $g', g^+, V, H$ bands; for the $z\sim5$ stacks, we excluded the $V_{606}$ and the $i_{775}$ bands; for the $z\sim6$ stack, we excluded the $i_{775}$ band while for the $z\sim7$ stack we excluded the $I_{814}$ band. The stacked SEDs together with the best-fit templates from FAST are presented in Figure \ref{fig:stack}. 

The photometric redshifts measured from the stacked SED are all consistent with the values of the corresponding redshift bin; the difference of the photometric redshifts of the median stacked and the median of the photometric redshifts of the individual sources in each subsample is $\Delta z/(1+z)\lesssim0.05$, i.e. within the uncertainties expected for photometric redshifts. 

The inset in the left panel of Figure \ref{fig:stack} presents the slope of the UV continuum ($\beta$), measured on the stacked photometry, as a function of absolute magnitude $M_{z'}$. The stacked SEDs at $z\sim4$ and $z\sim5$ are characterised by a trend in the UV slope, with bluer slopes for low-luminosity galaxies,  particularly evident for the $z\sim4$ stacks, and qualitatively consistent with the results of e.g., \citet{gonzalez2011} and \citet{oesch2013}. Our measurements do not present evidence for evolution of $\beta$ with redshift at fixed luminosity ($\approx M_*$), although the large uncertainties in $\beta$ (especially for the higher redshift bins) may be blurring trends. Furthermore, the SEDs in the rest-frame UV wavelength region of the 4-5 brightest $z\sim4$ stacks do not differ too much one from each other, while they differ substantially at wavelengths redder than the Balmer/4000\AA~break. 

\begin{figure}
\includegraphics[width=9cm]{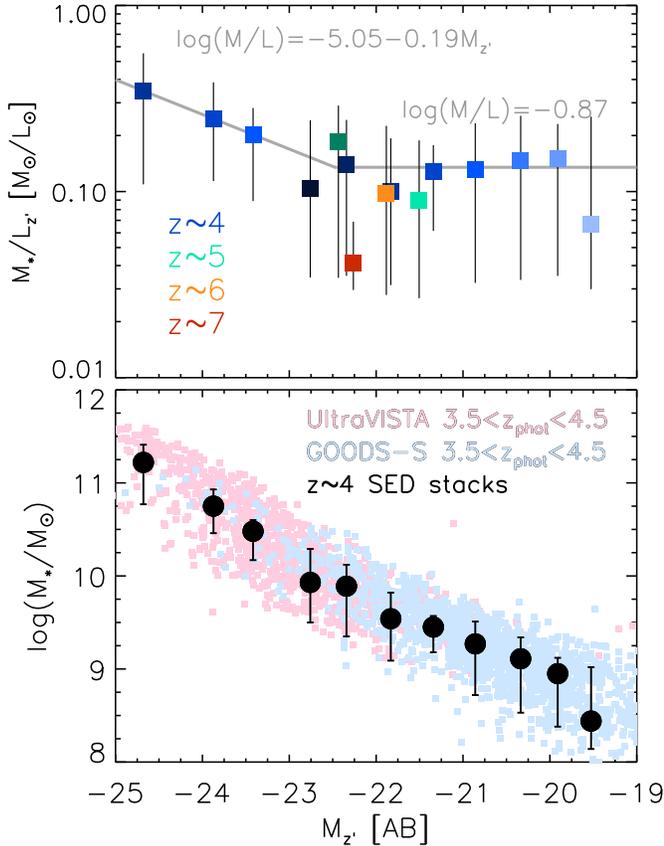}
\caption{{\bf Top panel:} The colored filled squares with error bars present the mass-to-light ratio ($M_*/L_{z'}$) from our stacking analysis as a function of absolute magnitude in the $z'$ band ($M_{z'}$). The color codes in the legend identify  the four redshift bins considered in this work. The assumed constant and best fit log-linear relations to the $z\sim4$ points are displayed by the grey lines and are expressed by the grey labels. {\bf Bottom panel:} Relation between the stellar mass and luminosity in the rest-frame $z'$ band, expressed in terms of absolute magnitude, obtained from samples selected through photometric redshifts (i.e., no LBG selection) over the COSMOS/UltraVISTA and CANDELS/GOODS-S fields (pink and light blue points, respectively). The black solid circles with errorbars mark our $z\sim4$ estimates of stellar mass and luminosity from the stacking analysis. Unlike the $M_{UV}-M_{z'}$ plane (Figure \ref{fig:uvopt}), the $M_*-M_{z'}$ follows a monotonic relation, with virtually no outliers over a wide range in luminosity and stellar mass. No significant offset is observed between our stacking measurements and those from the photo-z selected sample. \label{fig:ML}}
\end{figure}

Recently, \citet{oesch2013} presented stacked SEDs at $z\sim4$ in bins of $z$-band absolute magnitudes from a sample of galaxies based on CANDELS/GOODS-S, HUDF and HUDF09-2. The sample benefits from deep IRAC $3.6\mu$m and $4.5\mu$m imaging from the IRAC Ultra Deep Field (IUDF) program. The stacked SEDs (see e.g., their Fig. 2) show a clear trend of redder colors with increasing rest-frame $z$-band luminosity, in particular for $M_z<-21.5$. Our stacked SEDs are in qualitative agreement with those of \citet{oesch2013}, confirming the observed trend with luminosity. Furthermore, thanks to the wide area offered by UltraVISTA which provides coverage for even brighter sources, we are able to extend the trend to even more luminous galaxies.

In the top panel of Figure \ref{fig:ML} we present the $M_*/L_{z'}$ measured\footnote{We adopted $M_{z',\odot}=4.52$ AB} from the stacked SEDs as a function of $z'$ absolute magnitude. Total uncertainties were obtained by propagating the 68\% confidence intervals in stellar mass generated by FAST and the uncertainties in luminosity, taken as the flux uncertainties from stacking. At $z\sim4$ the $M_*/L_{z'}$ ratio is consistent with being constant for $M_{z'}$ fainter than $\sim-22.5$~mag. We find: 
\begin{equation} \label{eq:ML_const}
\log(M_*/L_{z'})=-0.87\pm0.09, \ \mathrm{for }\  M_{z'}\gtrsim -22.5
\end{equation}
For $M_{z'}\lesssim-22.5$~mag there is indication of $M_*/L_{z'}$  increasing with the luminosity, although the error bars are large. The best-fit SEDs of the most luminous stacks have a nearly constant age ($\sim10^{8.8}$ yr), and show an $A_V$ slightly increasing with stellar mass (from 1.0 to 1.2~mag). A linear fit of the $\log(M_*/L_{z'})$ values for $M_{z'}\lesssim-22.5$~mag resulted in the following relation: 
\begin{equation} \label{eq:ML}
\begin{split}
\log(M_*/L_{z'})=(-5.1\pm4.7) - & (0.19\pm0.18)\times M_{z'}, \\
&\mathrm{for }\  M_{z'}\lesssim -22.5
\end{split}
\end{equation}
Our linear relation recovers the constant value of $\log(M_*/L_{z'})=-0.87$ at $M_{z'}\sim-22.5$. However, the uncertainties on the fit parameters make the above relation also  consistent with a constant value. 

The $M_*/L_{z'}$ ratios for the $z\sim5$ and $z\sim6$ are consistent with $M_*/L_{z'}$ measurements of the $z\sim4$, $M_z>-22.5$~AB stacks. The $M_*/L_{z'}$ for the $z\sim7$ bin is consistent with the average $M_*/L_{z'}$ only at $\sim3 \sigma$ level. We note however, that the $z\sim7$ sample only includes two sources, therefore reducing the statistical significance of the observed disagreement.

The above results are consistent with what already observed in Figure \ref{fig:uvopt}. In Sect. \ref{sect:uvopt} we showed that galaxies more luminous than $M_{z'}\sim-22.5$~mag form a cloud in the rest-frame UV-$z'$ plane around $M_\mathrm{UV}\sim-21.4$~mag. From our stacking analysis, the average apparent magnitude at $\lambda_\mathrm{obs}\sim 8000$\AA~(i.e., the rest-frame UV$_{1600}$) for the stacked SEDs with $[4.5\mu\mathrm{m}]<23.5$~AB is $\sim24.7$~AB, which at $z\sim4$ corresponds to an absolute magnitude $M_{\mathrm{UV},1600}\sim-21.4$. According to the above relation, the stellar mass corresponding to $M_{z'}\sim-23$~mag is $\log(M_*/M_\odot)\sim10.3$. This behaviour warns about the potential biases that can occur when adopting the UV luminosity and $M_*/L_{UV}$ in the measurement of stellar masses, in particular for massive galaxies.

A constant  $M_*/L$ is equivalent to a slope of $-0.4$ in the log(stellar mass) - absolute magnitude plane. Our result at $z\sim4$, obtained for galaxies with $M_{z'}<-23$~mag, is consistent with the $\sim-0.4$ slopes found in the stellar mass - $M_{UV}$ plane (see e.g., \citealt{duncan2014,grazian2015}). Steeper slopes, as those found by \citet{stark2009,gonzalez2011,lee2011,mclure2011,song2016}, require the $\log(M_*/L)$ to decrease for fainter galaxies or increase for brighter galaxies or a combination of both effects. The origin for this is still unclear as it could be a mix between selection effects (see e.g. \citealt{grazian2015}) and nebular emission contamination which could boost the stellar masses of the more luminous galaxies (e.g., \citealt{song2016}).

In the bottom panel of Figure \ref{fig:ML} we compare the measurements of stellar mass and luminosity recovered from our stacking analysis to a \emph{control} sample. This sample is composed by individual measurements selected from the COSMOS/UltraVISTA and CANDELS/GOODS-S catalogs to have photometric redshifts $3.5<z_\mathrm{phot}<4.5$.

The individual measurements of the control sample distribute according to a monotonic relation defining a \emph{main sequence}, with a scatter of about $\sim0.7$~dex. This correlation holds over a wide range of values, both in stellar mass and in luminosity. In particular, we notice the absence of any turnover (as instead is observed when considering $M_\mathrm{UV}-M_*$ - see e.g., \citealt{spitler2014} or, equivalently, our Figure \ref{fig:uvopt}),  and virtually no measurements outside the main sequence. 

The relation between the stellar mass and luminosity recovered from the stacking analysis is in excellent agreement with the values of the control sample. We remark here that the control sample was selected based exclusively on photometric redshifts criteria. The agreement between our stacking results in the GOODS field, then, indicate that the stacking does not suffer any major bias from the LBG selection. This result is not unexpected, though, given the low fraction of objects missed by the LBG selection for stellar masses $\log(M_*/M_\odot)\lesssim10.2$, as we showed in Sect. \ref{sect:biases}.

Together, these two results increase our confidence on the reliability of the $z'$ band as a proxy for the stellar mass and on the robustness of our stacking analysis.

\subsection{Evolution of the $z'$-band Luminosity Function}

\begin{table}
\caption{$V_\mathrm{max}$ measurements of the Luminosity Functions  \label{tab:LF}}
\begin{tabular}{cccDr}
\hline
\hline
\decimals
$z$ & $M_{z'}$ & $\Delta M_{z'}$ & \multicolumn2c{$\Phi$} &\#  \\  
bin & (mag) & (mag) &  \multicolumn2c{($10^{-5}$ Mpc$^{-3}$ mag$^{-1}$ )} & gal. \\ 
\hline
$z\sim4$ & $-25.25$ & 0.25 & $0.044^{+0.102}_{-0.041}$ &    1 \\
 & $-25.00$ & 0.25 & $0.088^{+0.121}_{-0.068}$ &    2 \\
 & $-24.75$ & 0.25 & $0.22^{+0.18}_{-0.13}$ &    5 \\
 & $-24.50$ & 0.25 & $0.70^{+0.37}_{-0.35}$ &   16 \\
 & $-24.25$ & 0.25 & $1.27^{+0.62}_{-0.59}$ &   29 \\
 & $-24.00$ & 0.25 & $1.71^{+0.80}_{-0.78}$ &   39 \\
 & $-23.75$ & 0.25 & $2.4^{+ 1.1}_{- 1.1}$ &   55 \\
 & $-23.50$ & 0.25 & $3.9^{+ 1.8}_{- 1.7}$ &   78 \\
 & $-23.25$ & 0.25 & $6.6^{+ 3.0}_{- 2.9}$ &  100 \\
 & $-23.00$ & 0.25 & $10.8^{+ 4.8}_{- 4.8}$ &   86 \\
 & $-22.75$ & 0.25 & $16.8^{+ 7.6}_{- 7.5}$ &   64 \\
 & $-22.50$ & 0.25 & $25.^{+  11.}_{-  11.}$ &   72 \\
 & $-22.25$ & 0.25 & $28.6^{+ 7.5}_{- 7.2}$ &   64 \\
 & $-22.00$ & 0.25 & $44.^{+  11.}_{-  11.}$ &   87 \\
 & $-21.75$ & 0.25 & $62.^{+  15.}_{-  15.}$ &  123 \\
 & $-21.50$ & 0.25 & $65.^{+  16.}_{-  15.}$ &  128 \\
 & $-21.25$ & 0.25 & $95.^{+  22.}_{-  22.}$ &  186 \\
 & $-21.00$ & 0.25 & $101.^{+  24.}_{-  23.}$ &  194 \\
 & $-20.75$ & 0.25 & $103.^{+  24.}_{-  24.}$ &  196 \\
 & $-20.50$ & 0.25 & $126.^{+  29.}_{-  29.}$ &  207 \\
 & $-20.25$ & 0.25 & $162.^{+  38.}_{-  37.}$ &  193 \\
 & $-20.00$ & 0.25 & $191.^{+  46.}_{-  45.}$ &  119 \\
 & $-19.75$ & 0.25 & $215.^{+  59.}_{-  56.}$ &   49 \\
 &  &  & . & \\ 
$z\sim5$    & $-23.75$ & 0.50 & $0.56^{+0.76}_{-0.41}$ &    2 \\
 & $-23.25$ & 0.50 & $2.6^{+ 1.5}_{- 1.2}$ &    9 \\
 & $-22.75$ & 0.50 & $4.6^{+ 2.2}_{- 2.0}$ &   16 \\
 & $-22.25$ & 0.50 & $8.2^{+ 3.7}_{- 3.4}$ &   20 \\
 & $-21.75$ & 0.50 & $36.^{+  17.}_{-  15.}$ &   17 \\
 & $-21.25$ & 0.50 & $83.^{+  50.}_{-  41.}$ &    8 \\
 &  &  & . & \\
 $z\sim6$    & $-23.40$ & 0.80 & $0.20^{+0.46}_{-0.19}$ &    1 \\
 & $-22.50$ & 1.00 & $3.2^{+ 2.9}_{- 2.1}$ &    4 \\
 & $-21.65$ & 0.70 & $26.^{+  21.}_{-  16.}$ &    5 \\
 &  &  & . & \\
 $z\sim7$  & $-22.25$ & 0.50 & $2.9^{+ 4.1}_{- 2.4}$ &    2 \\
\hline
\end{tabular}
\end{table}

\begin{figure*}
\includegraphics[width=18cm]{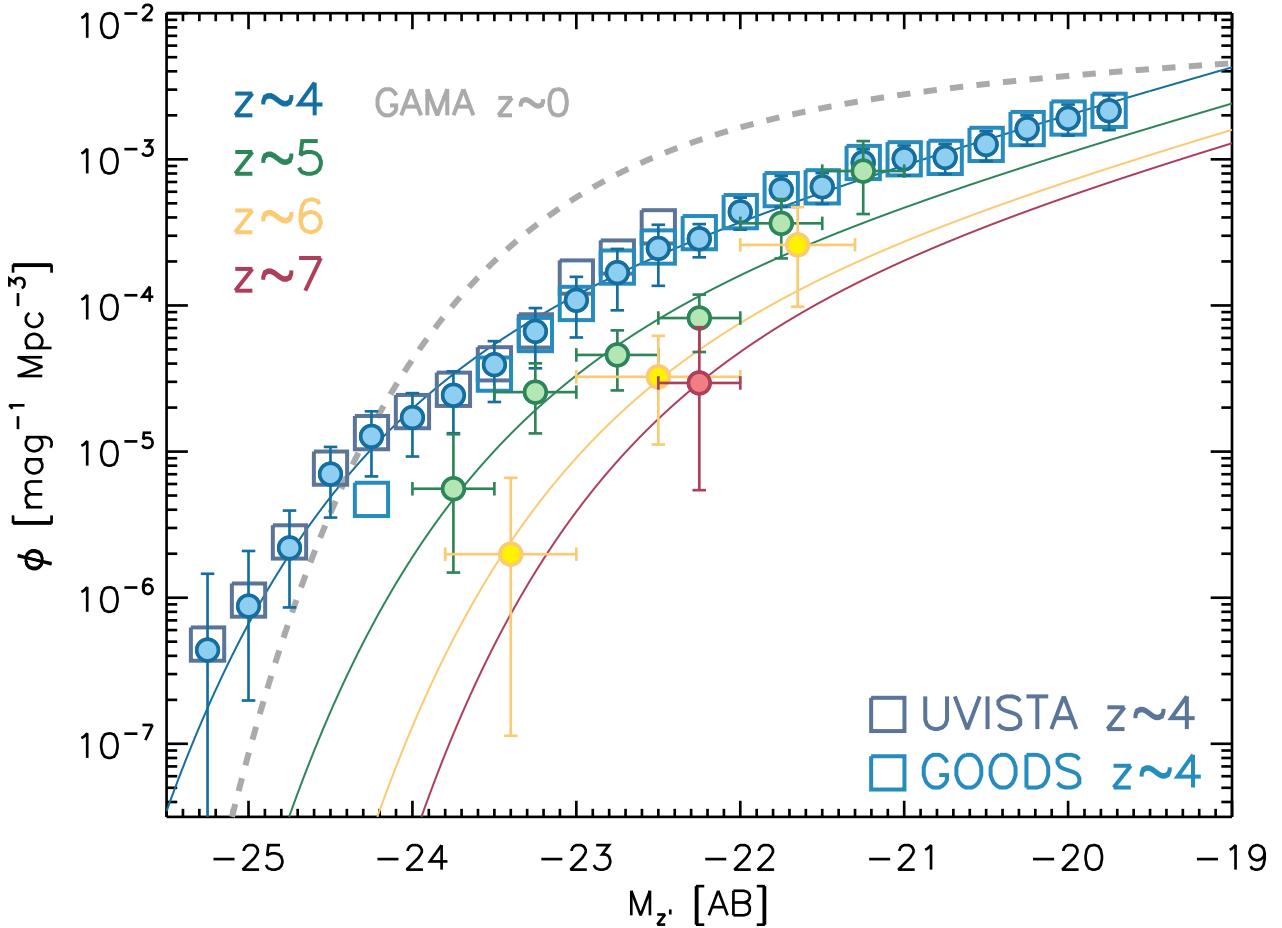}
\caption{The colored filled circles mark our measurements of the $1/V_\mathrm{max}$ LF in the four redshift bins as detailed by the legend in the top-left corner. Error bars include the contribution from Poisson noise and cosmic variance. For the $z\sim4$ bin we also present the individual LF from UltraVISTA (darker blue open squares) and GOODS-N/S (lighter blue open squares). These two latter measurements are consistent with each other where they overlap and with the LF from the composite $z\sim4$ sample. For ease of representation we omit the uncertainties of the individual UltraVISTA and GOODS-N/S LFs. The colored solid curves mark the best-fit Schechter functions at the corresponding redshift. The grey dashed line represents the $z\sim0$ LF from GAMA (\citealt{kelvin2014}). \label{fig:LF}}
\end{figure*}

The LFs were measured adopting the $1/V_\mathrm{max}$ estimator \citep{schmidt1968}. Although this method is intrinsically sensitive to local overdensities of galaxies,  at $z>4$ the clustering is expected to be negligible. On the other hand, the $1/V_\mathrm{max}$ method directly provides the normalisation of the LF. Furthermore, and most importantly, the \emph{coherent analysis} extension developed by \citet{avni1980} is key to this work. 

As we showed in Sect. \ref{sect:data_sets}, our composite sample is based on a dual-band flux selection, corresponding to a double flux threshold. The detection process introduces the first flux cut in the corresponding band ($K_\mathrm{s}$ or $\chi^2$ image built from the HST NIR bands, for the UltraVISTA and GOODS-N/S sample, respectively). The S/N cut on the flux in the IRAC band closest to the rest-frame $z'$ is responsible for the second flux threshold in the relevant IRAC band. 

For each galaxy in the sample, each flux threshold generates an upper limit to the redshift the specific  galaxy can have and still be included in the sample.  These different upper limits in redshift correspond to different comoving volumes for each object which could potentially enter the $V_\mathrm{max}$ computation. The coherent approach allowed us to take this double selection into consideration in a consistent way: the upper limit in redshift, used to compute the comoving volume, was taken to be the smaller one among the two redshift upper limits computed based on the threshold in the corresponding selection band.  Furthermore, as we showed in Figure \ref{fig:exptime_maps}, the depth of the IRAC mosaics is highly inhomogeneous. Therefore, for the computation of the comoving volumes in each field, we divided the IRAC footprint into a number of sub-fields, such that each sub-field was characterised by nearly homogeneous depths in both the detection\footnote{Since it is not straightforward to associate a limiting magnitude to a $\chi^2$ image from the combination of different filters, we considered the WFC3/$H_{160}$ the relevant band for the depth of the detection in the GOODS-N/S fields.} and in the relevant IRAC band. Again, the \citet{avni1980} prescription allowed us to analyse the different sub-samples coherently.

Comoving volumes were computed differently depending on the field and on the band driving the selection. For the galaxies in the GOODS-N/S fields, we used the comoving volumes computed by \citet{bouwens2015}. These volumes were estimated using an extensive Monte Carlo simulation based on real data. Sources were added to the different mosaics and recovered following the same procedure applied for the assembly of the LBG sample. Such volume estimates natively take into account the selection effects at the detection stage, correct for flux-boosting effects and contamination by lower redshift interlopers and brown dwarfs. The volumes $V_i$ for those objects $i$ in the GOODS-N/S fields whose redshift upper limit $z_\mathrm{up}$ was driven by the IRAC S/N threshold ($z_\mathrm{up}\equiv z_{\mathrm{up},\mathrm{IRAC}}$) were rescaled by the ratio between the volume associated to the redshift upper limit from the IRAC band $V_i({z_{\mathrm{up},\mathrm{IRAC}}})$ and the redshift upper limit in the $\chi^2$ image $V_i({z_{\mathrm{up},\chi^2}})$:
\begin{equation}
V_{i,\mathrm{IRAC}}=V_{i,\mathrm{GOODS}} \times \frac{V_i({z_{\mathrm{up},\mathrm{IRAC}}})}{V_i({z_{\mathrm{up},\chi^2}})}
\end{equation}
For the UltraVISTA sample, the volumes were computed directly from the limits in redshift corresponding to the flux limits in the $K_\mathrm{s}$ and $4.5\mu$m bands.

We computed the LF in four redshift bins centred at $z\sim4$, $z\sim5$, $z\sim6$ and $z\sim7$. Although the IRAC data potentially allowed us to consider galaxies at $z\sim8$, we did not find any candidate with reliable flux measurement in the IRAC $5.8\mu$m and $8.0\mu$m. Uncertainties on the LF measurements were derived by combining in quadrature the Poisson noise in the approximation of \citet{gehrels1986}  to an estimate of cosmic variance from the recipe of \citet{moster2011}. The average cosmic variance value obtained for the $z\sim4$ UltraVISTA sample was $\sim0.43$; the average cosmic variance estimates for the GOODS-N/S sample were $\sim0.27, \sim0.41, \sim0.58$ and $\sim0.80$, respectively for the $z\sim4, 5, 6$ and $z\sim7$ redshift bins. The high values of the cosmic variance  registered for all redshifts and luminosities are the dominant source of stochastic uncertainties in our LF measurements. 

Our LF measurements are presented in Figure \ref{fig:LF} and in Table \ref{tab:LF}.  The LF at $z\sim7$ consists of a single-bin measurement and is characterized by large uncertainties which do not allow us to properly constrain its shape. The absolute magnitude range of the $z\sim4$ LF spans $\sim5$ magnitudes, $\gtrsim3\times$ more than the magnitude range of the $z\ge5$ LFs.  The larger absolute magnitude range available at $z\sim4$ is the result of a number of distinct factors. First, the increased depth in the $4.5\mu$m band from the combination of \textit{Spitzer}/IRAC cryogenic and post-cryogenic epochs enables to reach fainter absolute magnitudes than the cryogenic only $5.8+8.0\mu$m data at $z\ge 5$. Second, the smaller PSF size of the $4.5\mu$m data compared  to the $5.8\mu$m and $8.0\mu$m bands allows to reach fainter fluxes for the same exposure time and detector efficiency.  Third, the availability of COSMOS/UltraVISTA data over an area $\sim4\times$ larger than the GOODS-N/S footprint allowed us to recover the exponential decline of the bright end of the $z\sim4$ LF, otherwise inaccessible by the small footprint of the GOODS-N/S mosaics.

\begin{figure*}
\includegraphics[width=16cm]{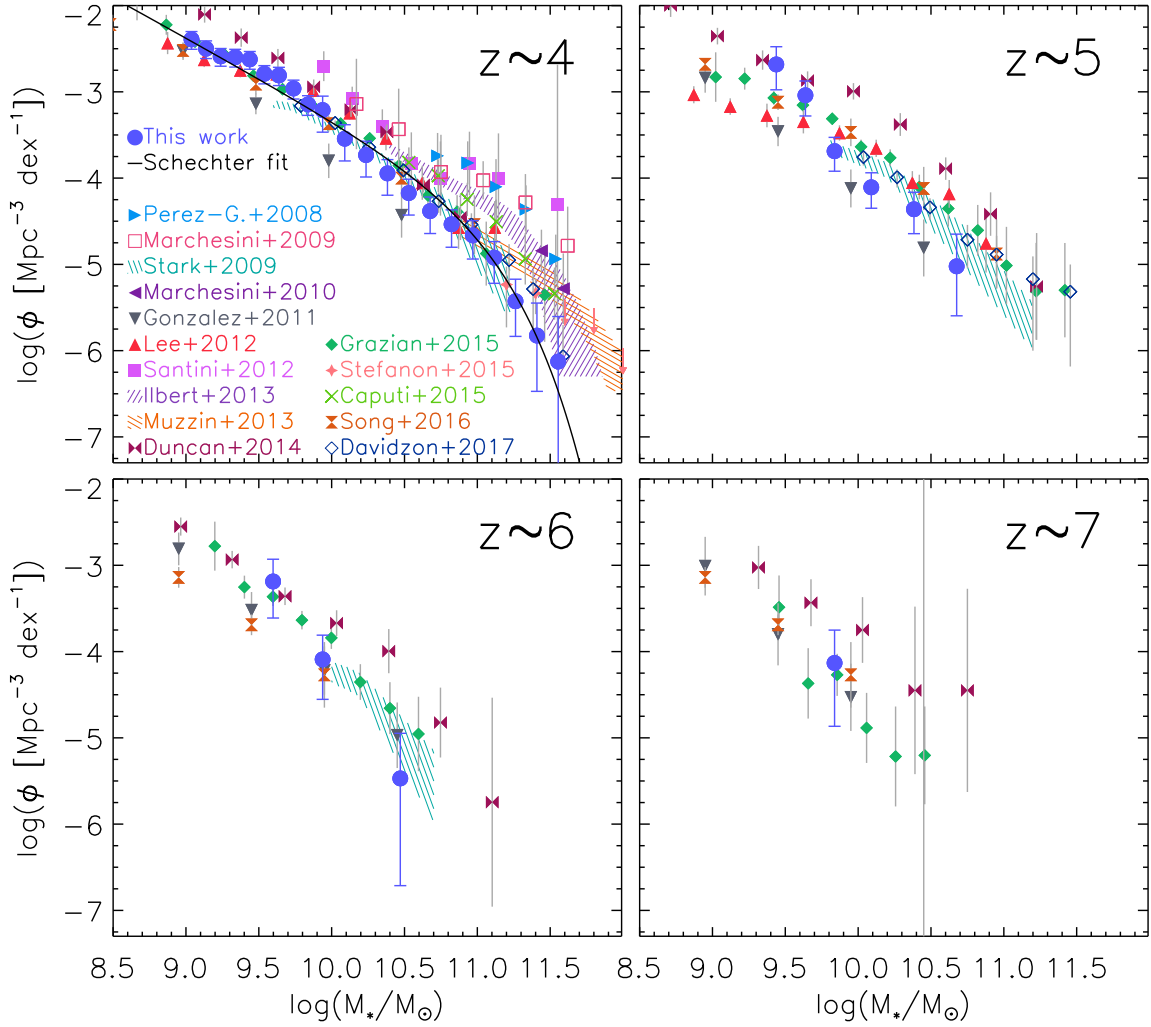}
\caption{Our estimates for the SMF are marked by the filled blue circle with error bars. Top to bottom, left to right the panels present the SMF at $z\sim4, 5, 6$ and $7$, respectively. Measurements of the SMF from the literature are also shows, following the symbols of the legend in the top-left panel. The same plotting conventions are applied to all panels. Our SMFs are in good agreement with previous determinations. \label{fig:SMF}}
\end{figure*}

In order to verify the consistency of the $z\sim4$ LF with respect to the GOODS-N/S and UltraVISTA data, we also computed the LF separately on each one of these two datasets. The resulting LFs are marked  in Figure~\ref{fig:LF} by the open squares and show a good agreement with the LF from the composite sample.

The large uncertainties associated to the number density measurements at $z\sim5,6,7$ do not allow us to disentangle whether the evolution is in luminosity, in number density or in both. In Sect. \ref{sect:lf_schechter} we attempt to analyse this in a more quantitative way.

The dashed grey curve in Figure \ref{fig:LF} marks the $z\sim0$ LF of \citet{kelvin2014} measured with data from the Galaxy and Mass Assembly (GAMA) survey (\citealt{driver2009}). Compared to our lowest redshift LF measurement, the $z\sim0$ LF is characterised by a steeper decay for $M_{z'}\lesssim-24.5$. Fully understanding the evolution of the LF from $z\sim4$ to $z\sim0$ goes beyond the scope of the present work.  However, qualitatively, a decrease in luminosity at $z\sim0$ compared to $z\sim4$ is expected, considering the lower values of the star-formation rate density at $z\sim0$ than at $z\sim4$ (e.g. \citealt{madau2014}) and that the $z'$ band may retain the effects of the recent star formation history. This is particularly true for the bright-end of the LF: indeed, the brightest ($\sim$ most massive) galaxies that are still forming stars at $z\sim4$ are likely to become quenched by $z\sim0$ (e.g., \citealt{muzzin2013b}).

\subsection{Evolution of the Stellar Mass Function}

\label{sect:smf}

\begin{table}
\caption{Stellar mass function measurements \label{tab:SMF}}
\begin{tabular}{cDD}
\hline
\hline
\decimals
$z$ &\multicolumn2c{Stellar mass}  & \multicolumn2c{$\Phi$}   \\  
bin & \multicolumn2c{$\log(M_*/M_\odot)$} &  \multicolumn2c{($10^{-5}$ Mpc$^{-3}$ dex$^{-1}$ )}  \\ 
\hline
$z\sim4$   &   11.55 & $0.075^{+0.174}_{-0.070}$  \\
 &   11.41 & $0.15^{+0.21}_{-0.12}$  \\
 &   11.26 & $0.37^{+0.30}_{-0.23}$  \\
 &   11.12 & $1.20^{+0.64}_{-0.59}$  \\
 &   10.97 & $2.2^{+ 1.0}_{- 1.0}$  \\
 &   10.82 & $2.9^{+ 1.4}_{- 1.3}$  \\
 &   10.68 & $4.1^{+ 1.9}_{- 1.9}$  \\
 &   10.53 & $6.7^{+ 3.0}_{- 3.0}$  \\
 &   10.38 & $11.3^{+ 5.0}_{- 5.0}$  \\
 &   10.24 & $18.5^{+ 8.3}_{- 8.2}$  \\
 &   10.09 & $29.^{+  13.}_{-  13.}$  \\
 &    9.94 & $61.^{+  28.}_{-  27.}$  \\
 &    9.84 & $71.^{+  19.}_{-  18.}$  \\
 &    9.74 & $109.^{+  27.}_{-  27.}$  \\
 &    9.64 & $155.^{+  37.}_{-  37.}$  \\
 &    9.54 & $162.^{+  39.}_{-  38.}$  \\
 &    9.44 & $238.^{+  56.}_{-  55.}$  \\
 &    9.34 & $252.^{+  59.}_{-  58.}$  \\
 &    9.24 & $258.^{+  60.}_{-  60.}$  \\
 &    9.14 & $316.^{+  73.}_{-  73.}$  \\
 &    9.04 & $405.^{+  94.}_{-  94.}$  \\
 &    8.94 & $477.^{+  115.}_{-  114.}$  \\
 &    8.84 & $537.^{+  147.}_{-  141.}$  \\
 & . & .  \\
$z\sim5$    &   10.68 & $0.95^{+1.29}_{-0.70}$  \\
 &   10.38 & $4.4^{+ 2.5}_{- 2.1}$  \\
 &   10.09 & $7.8^{+ 3.7}_{- 3.4}$  \\
 &    9.84 & $20.5^{+ 9.2}_{- 8.5}$  \\
 &    9.64 & $91.^{+  42.}_{-  39.}$  \\
 &    9.44 & $208.^{+  125.}_{-  102.}$  \\
 & . & .  \\
$z\sim6$    &   10.47 & $0.34^{+0.79}_{-0.32}$  \\
 &    9.94 & $8.1^{+ 7.4}_{- 5.3}$  \\
 &    9.60 & $65.^{+  53.}_{-  40.}$  \\
 & . & .  \\
$z\sim7$   &    9.84 & $7.4^{+10.4}_{- 6.0}$  \\
\hline
\end{tabular}
\end{table}

We generated SMF measurements taking advantage from  the mass-to-light ratios we measured in Section \ref{sect:ml_ratios}. The feature that the $M_*/L_{z'}$ does not decrease with luminosity makes the shape of the LF to resemble that of the SMF, allowing us to attempt a simple and straightforward conversion of the LF into the SMF. Other $M/L$ relations allow for the recovery the SMF from the LF, although in a less straightforward way. We further discuss this in Sect. \ref{sect:lf_vs_smf}.

We adopted the following very simple procedure. We assumed that the constant $\log(M_*/L_{z'})$ and the linear relation observed at $z\sim4$ (Eq. \ref{eq:ML_const} and \ref{eq:ML}) were valid at all redshifts.  The absolute magnitudes corresponding to the bin centers of the LFs were converted into stellar mass applying the relevant $\log(M_*/L_{z'})$ relation depending on the $M_{z'}$ value (see Eq. \ref{eq:ML_const} and Eq. \ref{eq:ML}). We then differentiated the two relations, solving for $dM_*$. The obtained values, specific for each $M_*$ bin, were used to rescale the LF normalisation, to take into account the change in units from mag$^{-1}$ to dex$^{-1}$.

Our SMF measurements are presented in Figure \ref{fig:SMF} and Table \ref{tab:SMF}.  Unsurprisingly, the $z\sim4$ SMF covers a range in stellar mass wider than the $z\sim5,6,7$ SMFs, for the same reasons we described for the LF.

In Figure \ref{fig:SMF} we also plot a compilation of SMF measurements  from the literature (\citealt{perez-gonzalez2008, marchesini2009, stark2009, marchesini2010,gonzalez2011, lee2012, santini2012, muzzin2013b, ilbert2013, duncan2014, stefanon2015, caputi2015, grazian2015, song2016,davidzon2017}). At $z\sim4$, starting from the low-mass end where the measurements are generally quite consistent with each other, the discrepancies increase with increasing stellar mass. One possible reason for the increased dispersion at higher masses is that galaxies constituting the low-mass end are mostly star-forming. Their redshift can then be assessed through the location of the observed Lyman break (either from dropouts or photometric redshift selections). The massive end, instead, possibly also includes more evolved and/or dusty systems and it is therefore more sensitive to the degeneracy in identifying the observed break as either the Balmer/4000\AA~break or the Lyman break. 

Our $z\sim4$ SMF determination is in good agreement with the SMFs of \citet{stark2009}, \citet{lee2011}, \citet{stefanon2015}, \citet{caputi2015}, \citet{grazian2015}, \citet{song2016} and \citet{davidzon2017}. This is quite remarkable, since these SMFs have been recovered from different selection techniques. Specifically, \citet{stark2009} and \citet{lee2011} measurements are based on dropouts samples from \emph{HST}/WFC3 data; the SMF of \citet{grazian2015} was built from a $H_{160}$-detected  photometric redshift sample over the  CANDELS/GOODS-S and CANDELS/UDS fields; \citet{stefanon2015} assembled a composite sample complementing a $K_\mathrm{s}$-detected catalog from UltraVISTA data with detections in IRAC $3.6\mu$m and $4.5\mu$m bands; \citet{caputi2015} measurements are based on a $K_\mathrm{s}$-detected SMF complemented by SMF measurements from detections in IRAC $4.5\mu$m. The sample selection of both \citet{stefanon2015} and \citet{caputi2015} relies on photometric-redshift measurements. \citet{song2016} recovered the SMF measurements converting the UV LF into SMF through a linear $M_*-M_\mathrm{UV}$ relation for $M_*<10^{10}M_\odot$, complemented by boostrapped estimates at $M_*>10^{10}M_\odot$. Finally, the SMF of \citet{davidzon2017} were based on a photometric-redshift sample from $K_\mathrm{s}$-detection in COSMOS/UltraVISTA DR2 mosaics. On the other side, the normalisation of our SMF is higher than \citet{gonzalez2011} SMFs; these measurements were obtained by converting the observed UV LF into SMF through $M_*/L_\mathrm{UV}$ measurements. The discrepancy with our SMF (and the bulk of the other SMF determinations) could be due to a steeper $M_*/L_\mathrm{UV}$ relation found by \citet{gonzalez2011} and consequent lower normalisation term. 

At the massive end ($\log(M_*/M_\odot) \gtrsim 11.2$), we observe a discrepancy between our $z\sim4$ SMF and {some of } the corresponding measurements from the literature (e.g., \citealt{muzzin2013b, ilbert2013}). This discrepancy could, at least in part, be explained by our SMF lacking any scatter in $M_*/L_{z'}$ for a given $L_{z'}$. Our stacking analysis, by construction, recovers median $M_*/L_{z'}$ ratios, potentially excluding extreme cases such as very dusty/old systems with very high stellar masses. However, the bottom panel of Figure \ref{fig:uvopt} and our discussion in Section \ref{sect:biases} show that our sample is not strongly biased against this class of objects in the limits of current data. Nonetheless, they do not allow us to properly ascertain their existence at lower stellar masses. Besides, because of the Eddington bias, a distribution in the observed $M_*/L_{z'}$ values for a specific luminosity would introduce a higher fraction of lower stellar mass objects scattered to higher stellar masses than the opposite, increasing the number density of the massive objects.  One additional potential reason for this discrepancy are differences between the current and previous estimates of photometric redshifts. Compared to \citet{muzzin2013b} or \citet{ilbert2013}, the DR2  version of the UltraVISTA catalog benefits from deeper NIR and IRAC data, providing improved photometric redshift constraints. It is noteworthy the agreement between our estimates and those of \citet{grazian2015} and \citet{davidzon2017}, both obtained from photometric redshift samples, suggesting that our composite sample suffers from small selection bias, even at the higher masses.

At $z\sim5$, the measurements of the low-mass end  ($\log(M_*/M_\odot)\sim9-9.5$) of the SMF are characterised by a larger scatter than for $\log(M_*/M_\odot)\sim10-10.5$. At higher stellar masses measurements are broadly consistent with each other mainly because of the large uncertainties on the number densities. Our SMF determination overlaps with the measurements of \citet{stark2009},\citet{gonzalez2011}, \citet{lee2011}, \citet{grazian2015}, \citet{song2016}  and \citet{davidzon2017}. However, it lies below the SMF measurements of \citet{duncan2014}.

At even higher redshift, the Poisson uncertainties on the number densities start to be of the same order as those from cosmic variance. The small number of galaxies at $z\sim6$ and $z\sim7$ (10 and 2 galaxies, respectively) generates large Poisson uncertainties which result in a broad agreement among the different SMF determinations. Our SMF overlaps with the measurements of \citet{stark2009}, \citet{gonzalez2011}, \citet{grazian2015} and \citet{song2016}. The large uncertainties make our SMF measurements roughly consistent also with those of \citet{duncan2014}.

\subsection{Schechter best-fit to the LF and SMF}
\label{sect:lf_schechter}

The shape of the LF and SMF of galaxies at high redshift is usually well described by a \citet{schechter1976} function over a wide range of redshifts ($2\lesssim z\lesssim10$, but see also e.g. \citealt{bowler2014}). We therefore fitted our LF measurements with a Schechter function:
\begin{equation}
\Phi(M)=\Phi^*\frac{\ln(10)}{2.5}10^{-0.4(M-\mathcal{M}^*)(\alpha+1)}\exp{(-10^{-0.4(M-\mathcal{M}^*)})}
\end{equation}
where $\mathcal{M}^*$ is the characteristic magnitude, corresponding to the knee of the density distribution, $\alpha$ is the faint end slope and $\Phi^*$ is a global normalisation factor. The single-bin measurement at $z\sim7$ does not allow us to properly inspect the shape of the LF. However, LF estimates at similar redshift, although in different bands, show that it is reasonably well determined and consistent with a single Schechter form (\citealt{bouwens2015, finkelstein2015a}) or with a double power-law \citep{bowler2014,bowler2017}.

We performed a best-fit to our LFs measurements using the Levenberg-Marquardt method. For the $z\sim4$ LF we left  the three Schechter parameters free to vary.  Since the faint-end slope of the $z\ge5$ LFs is poorly sampled, the best fits to the $z\ge5$ LFs were done fixing $\alpha$ to the value obtained for the $z\sim4$ LF.  The fit to the $z\sim5$ and $z\sim6$ LFs were done in three different configurations: both $\mathcal{M}^*$ or $\Phi^*$ as free parameters, with $\mathcal{M}^*$  as unique free parameter and with $\Phi^*$ as the only free parameter. We did not fit the $z\sim7$ LF leaving both  $\mathcal{M}^*$ or $\Phi^*$ as free parameters. The resulting best-fit Schechter functions for the case of pure luminosity evolution are marked in Figure \ref{fig:LF} with solid curves, while Table~\ref{tab:schechter} lists the recovered parameters and their uncertainties.

\begin{table}
\caption{Luminosity function best-fit Schechter parameters\label{tab:schechter}}
\begin{tabular}{cccc}
\hline
\hline
Redshift & $\Phi^*$ & $\mathcal{M}^*$ & $\alpha$ \\
bin & ($10^{-5}$ Mpc$^{-3}$ mag$^{-1}$) & (mag) & \\
\hline \\
4 & $19.4\pm7.1$ & $-23.38\pm0.19$ & $-1.79\pm0.09$ \\
   &            &      &      \\
5 &  $9.1\pm7.7$ & $-22.99\pm0.51$ & $-1.79$ \\
   &  $19.4$       & $-22.62\pm0.12$ & $-1.79$ \\
   &  $6.85\pm1.36$ & $-23.38$       & $-1.79$ \\ 
   &            &      &      \\
6 &  $51.1\pm36.0$ & $-21.77\pm0.42$ & $-1.79$ \\
  &  $19.4$       & $-22.09\pm0.17$ & $-1.79$ \\
   &  $8.59\pm5.40$ & $-23.38$       & $-1.79$ \\ 
   &            &      &      \\
7 &  $19.4$       & $-21.82\pm0.39$ & $-1.79$ \\
   &  $1.99\pm1.62$ & $-23.38$       & $-1.79$ \\ 
 
\end{tabular}
\end{table}

Visual inspection of the best-fitting Schechter function showed that, overall, there is a preference for a pure luminosity evolution against a pure density evolution. Moreover, when both $\mathcal{M}^*$ and $\Phi^*$ were left free to vary, the values of $\Phi^*$ were characterized by large uncertainties, making them consistent with the value of $\Phi^*$ at $z\sim4$, i.e., corresponding to no evolution with redshift. On the contrary, the values of $\mathcal{M}^*$ showed a more clear trend with redshift, increasing our confidence on the luminosity evolution. We note however, that these results potentially suffer from the limited coverage of $\mathcal{M}^*$. 

The luminosity evolution registered here is in contrast with the most recent measurements of the evolution of the UV LF at $z>4$, where indication is found that the characteristic magnitude evolves very little with redshift (likely constrained by the impact of dust extinction at high masses - \citealt{bouwens2009, reddy2010}), as most of the evolution seems to be driven by variation in the overall density (see e.g., \citealt{vanderburg2010, bouwens2015, finkelstein2015a}). 

The LFs presented in our work are the first determination of the rest-frame optical LF at $z\gtrsim4$ and therefore direct comparisons to previous estimates are not possible. However, recent works have recovered the measurement of the characteristic magnitude in the rest-frame $z'$-band at $z\sim4$ (\citealt{oesch2013}) or have studied the evolution of the LF up to $z\lesssim4$ in rest-frame optical bands close to $z'$ (\citealt{marchesini2012, stefanon2013}). In the following paragraphs we will compare our determination of the $z\sim4$ characteristic magnitude to the estimates from the above three works.

\citet{oesch2013} estimated the characteristic magnitude of the $z\sim4$ LF applying a correction based on the average $i_\mathrm{775}-[4.5]$ color to the characteristic magnitude of the $z\sim4$ UV LF. The obtained value, $\mathcal{M}^*_{z}=-21.7$~AB, is $\sim1.7$~mag fainter than what we found in this work. A possible reason for this large discrepancy is the lack of galaxies brighter than $M_{z}<-23$ from the sample of \citet{oesch2013}, which instead have become accessible through the deep and wide area UltraVISTA data.

\citet{marchesini2012} presented estimates of the rest-frame $V$-band LF at $z\lesssim4$, obtained from a composite sample including wide-area data from the NMBS, FIRES, FIREWORKS, HDFN, HUDF and GOODS-S programs. The characteristic magnitude recovered from a maximum likelihood analysis is $\mathcal{M}^*_V=-22.76^{+0.40}_{-0.63}$. This value is brighter than that found by \citet{oesch2013}, although it is still $\sim0.6$~mag fainter than the estimate in our work; nonetheless it is consistent at $1 \sigma$ with our estimate, considering the associated uncertainties. Our stacking analysis (see Figure \ref{fig:stack}) showed that the brighter galaxies have redder $(V-z')$ colors, reaching $(V-z')\sim1$~mag for the brightest stacks. Using our stacked SED at $M_{z'}\sim-23.5$, close to the value of the characteristic magnitude of our $z\sim4$ LF, we find a rest-frame color $(V-z')=+0.53$~mag, which, applied to the $\mathcal{M}^{*}_V$, gives a $\mathcal{M}^{*}_{z'}=-23.29$~mag, very close to our best-fit $\mathcal{M}^{*}_{z'}=-23.38$~mag.

\citet{stefanon2013} measured the evolution of the rest-frame $J-$ and $H-$band LF up to $z\sim3.5$ using a composite sample of galaxies from the MUSYC, FIRES and GOODS-CDFS programs. From maximum likelihood analysis, the characteristic magnitude of the $z\sim3.25$ rest-frame $J-$band LF was estimated to be $\mathcal{M}^{*}_{J}=-23.28^{+0.33}_{-0.29}$~mag. Applying the same analysis adopted for the comparison to \citet{marchesini2012}, we find a rest-frame color $(z'-J)=+0.11$~mag, and a corresponding $\mathcal{M}^{*}_{z'}=-23.17$~mag, still consistent with our estimate at $1\sigma$ level.\\

We also performed a Schechter fit to the SMFs. However, the differential $M_*/L_{z'}$ we measured at $z\sim4$ and applied to the $z\ge5$ LFs has the effect of \emph{stretching} the original LFs. The consequence of this stretching, together with the limited number of measurements available at each redshift, is that the $z\ge5$ SMFs do not show any robust evidence for the exponential cut at the massive end, preventing any reliable estimate of the characteristic stellar mass. A similar result was found by \citet{grazian2015}, even though the issue is not yet settled (see e.g., \citealt{caputi2015}). Therefore, we performed the Schechter fit only on the $z\sim4$ SMF, and obtained the following results: $\alpha=-1.93 \pm0.24$, $\Phi_*^*=(2.72\pm1.81)\times10^{-5}$~Mpc$^{-3}$~dex$^{-1}$ and $\log(M_*^*/M_\odot)=10.96\pm0.33$. The corresponding best-fit Schechter function is presented in Figure \ref{fig:SMF}.

The large uncertainties associated to the Schechter parameters make them consistent with most of the measurements from the literature. The low number of massive galaxies from the exponential part of the Schechter function, at the massive end, suffer from high uncertainties from both (relative) poisson noise and cosmic variance. The massive end of our SMF is based on the UltraVISTA sample, suggesting a more adequate coverage of massive galaxies.

\begin{figure*}
\includegraphics[width=16cm]{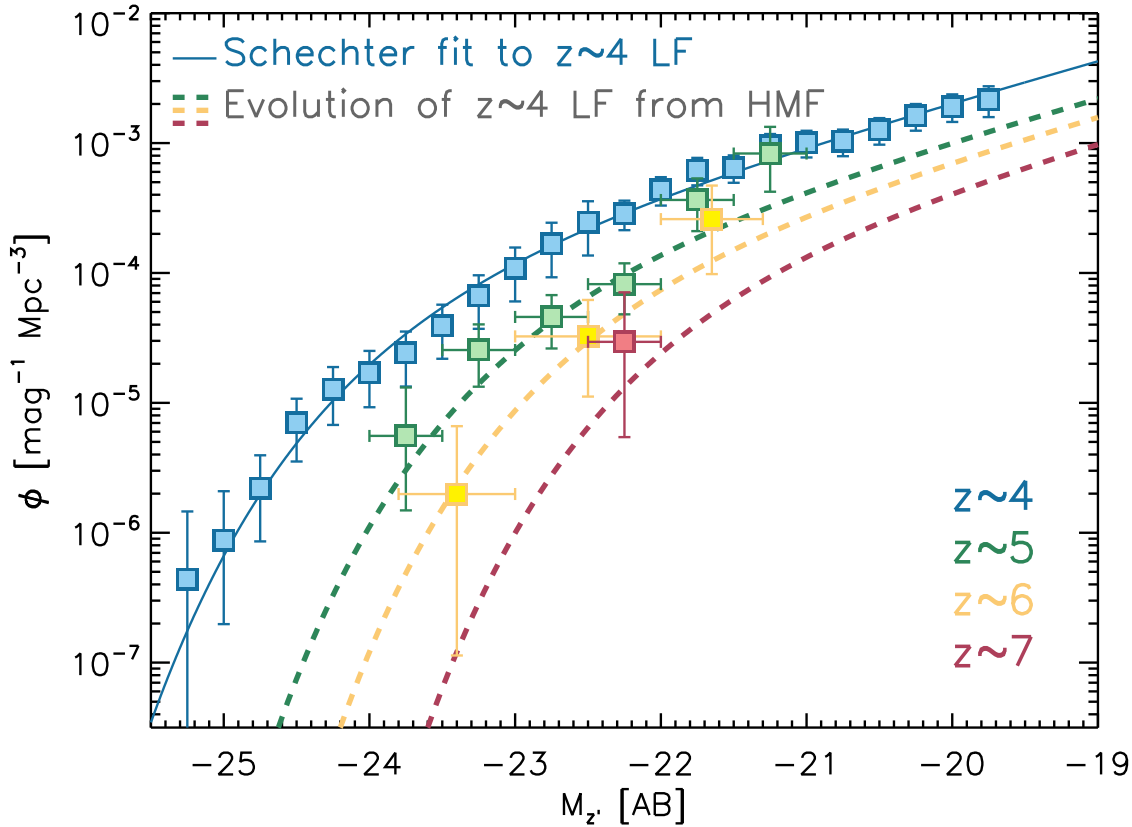}
\caption{The filled coloured squares mark our measurements of the $1/V_\mathrm{max}$ LF in the four redshift bins as detailed by the legend in the bottom-right corner. Error bars include the contribution from Poisson noise and cosmic variance. The solid blue curve marks the best-fit Schechter function at $z\sim4$, while the dashed curves present the $z\sim4$ Schechter function evolved in luminosity following the evolution in mass of the halo mass function relative to the $z\sim4$ HMF. \label{fig:LF4HMF}}
\end{figure*}

\section{Discussion}
\label{sect:halo}

\subsection{Rest-frame optical LF vs.  SMF measurements}
\label{sect:lf_vs_smf}

One of the main aim of this paper was to test the LF in rest-frame optical bands as a proxy for SMF measurements at high redshifts ($z\gtrsim4$). The compilation of SMF measurements presented in Figure \ref{fig:SMF} shows that for $z\ge4$ they are characterised by a scatter which can be as large as $\gtrsim1$~dex. This has immediate consequences on our understanding of more fundamental and global properties, like the evolution with cosmic time of the stellar mass density or the stellar-to-halo mass relation. The likely main reason for this large scatter is sample selection. In addition to this, further systematics may arise from our limited knowledge on some of the specific aspect characterising the stellar population of each galaxy, including contamination from emission lines (see e.g., \citealt{conroy2009,stark2009,behroozi2010}).

Reliable stellar mass measurements require coverage of rest-frame optical/NIR wavelengths, probed by  \emph{Spitzer}/IRAC at $z\gtrsim4$. In particular, current measurements of the low-mass end of the $z>4$ SMFs rely on IRAC data over the GOODS-N/S fields as these are the only fields providing  photometric coverage over a sufficient area and with the sufficient depth. The $3.6\mu$m and $4.5\mu$m bands, although characterised by deeper data than the $5.8\mu$m and $8.0\mu$m bands, at $z \gtrsim 4$ are potentially contaminated by nebular lines. Previous work has attempted to estimate the impact that nebular emission can have on the recovery of stellar mass, yet no concordance has been found so far. Specifically, estimates based on relations between the EW of $H\beta$ and the ionizing properties of the best-fit SEDs generally predict systematics $\lesssim 0.2$~dex and nearly independent on redshift (e.g., \citealt{duncan2014, grazian2015, salmon2015}). On the other side, estimates based on the measurements of the evolution of the $H\alpha$ EW with redshift indicate an increasing contribution of nebular lines in the stellar mass measurements, ranging from $\sim0$ dex at $z\sim4$ to $z\sim0.6$~dex at $z\sim7$ (e.g., \citealt{stark2013} who presented, among other estimates, an extrapolation to $z\sim 7$, later supported by observations - e.g. \citealt{smit2014, rasappu2016, faisst2016}). Such systematics in the measurement of stellar mass introduces up to $\sim0.5$~dex offset in the number densities for the higher redshift bins. To further complicate the picture, recent works have shown that when the correction for nebular line contamination is applied on a statistical basis, irrespective of the specific SEDs, it can even \emph{boost} the stellar mass measurements by up to $\sim 0.2-0.3$~dex (\citealt{stefanon2015, stefanon2017, nayyeri2017}).

One possible way for circumventing this problem is estimating the stellar mass exclusively from the $5.8\mu$m and $8.0\mu$m bands. These bands cover a region of the SEDs of $z\gtrsim4$ galaxies free from contamination by strong nebular lines. However, the larger PSF FWHM of \emph{Spitzer}/IRAC $5.8\mu$m and $8.0\mu$m bands compared to the $3.6\mu$m and $4.5\mu$m bands may introduce blending effects in the flux measurements. \citet{labbe2015} have shown that \texttt{mophongo}, the software we adopted to measure the fluxes in the IRAC bands, does not introduce any substantial systematics in the flux measurement, even in very crowded regions. Nonetheless, given the larger FWHM and the lower S/N characterizing the $5.8\mu$m and $8.0\mu$m data, one could expect an increased scatter in the flux measurements compared to the bluer IRAC bands.

Through the $M_*/L_{z'}$ we derived in the present work, the S/N cuts we applied to the flux in the IRAC $5.8\mu$m and $8.0\mu$m bands identify a range in stellar mass where their measurements can be considered reliable. The LF and SMF from this work, then, can be regarded as a indicative of most of the current SMF measurements at $z\ge 4$, as they are based on the subsample of objects with the highest S/N measurements in those bands more sensitive to the stellar mass and with reduced contamination from nebular emission. This is visible in Figure~\ref{fig:SMF}: at $z\sim5$ and above the lowest stellar mass over which our SMFs are defined is $\approx5-10\times$  higher than most current SMF determinations. Specifically, this also means that stellar masses below our low-mass limits are necessarily based on either very low S/N measurements in the IRAC $5.8\mu$m and $8.0\mu$m bands or on (still uncertain) correction for nebular emission contamination, or a combination of the two.

Our analysis showed that  SMFs consistent with the average determinations from the literature could be recovered by applying a simple $M_*/L_{z'}$ relation to the observed $z'$-band LF, with stellar masses measured from common stellar population parameters (e.g., delayed exponential SFH, solar metallicity, Chabrier IMF). Our simple transformation of the LF into SMF was supported by the non-decreasing $M_*/L_{z'}$ ratio for increasing luminosities observed at $z\sim4$. Relations between the $M_*/L$ and $L$ other than that (e.g., if the $M_*/L$ presented a minimum for some value of $L$) would still allow the conversion, but would require to consider, for specific bins of $M_*$, the contributions to the density originating from different bins of $L$. A non-decreasing $M_*/L_{z'}$, instead, constitutes an injective mapping between luminosity and stellar mass: galaxies with higher luminosity will always have higher stellar mass, and galaxies with lower luminosity will always have lower stellar mass.

A different scenario arises when multiple $M_*/L$ exist in correspondence to a single value of $L$, as it can be the case, for instance, of UV luminosity vs. stellar mass. Figure \ref{fig:uvopt} (together with the bottom panel of Figure \ref{fig:ML}) shows that for $M_\mathrm{UV} \gtrsim -20$~mag there are broadly two very different $M_*/L_\mathrm{UV}$ values. Specifically, this means that, if galaxies with the higher $M_*/L_\mathrm{UV}$ ratio are included in the sample, those bins of luminosity will include (and mix) the contribution from both high- and low-mass galaxies. If this effect is not properly taken into account, it introduces an over-estimate of the low-mass-end slope and an under-estimate of the massive end of the SMF. The only way to deal with this problem is to directly count the number of galaxies in each of the two $M_*/L$ bin.

Finally, the ideal rest-frame band for this kind of studies is probably one for which the $M_*/L$ ratio would not depend on the luminosity, as any luminosity dependence could potentially hide effects from e.g., SFH. Our measurements of the $M_*/L_{z'}$ relation suggest that, for a wide range in luminosity, they are consistent with a constant value. A log-linear relation arises at the bright- (massive-) end of the LF (SMF), suggesting that at these luminosities the $z'$-band holds at some level the signature of the stellar population age and/or of the dust content.

\subsection{Tracking the assembly of DM halos through the evolution of the rest-frame optical LF}
\label{sect:DM_assembly}

Given the potential systematics on the SMF measurements discussed above, LF estimates can provide a valid alternative for recovering the halo masses ($M_h$) for high-redshift galaxies through abundance matching techniques (e.g., \citealt{behroozi2013, finkelstein2015b, steinhardt2016}). To date, measurements of $z\ge4$ LFs are mostly available in the rest-frame UV. The adoption of UV LFs in the $M_h/L$ estimates provides information on the relative importance of star formation processes  (e.g., gas cooling, stellar ejections, SFR timescales) versus the hierarchical growth of the dark matter halos (\citealt{bouwens2015}). However, stellar masses are likely to be more strongly correlated with the halo masses than UV luminosities. 

The Schechter fits performed in Sect.~\ref{sect:lf_schechter} suggest that the evolution of the $z'$-band LFs can be accounted for by an increase of luminosity with cosmic time uniformly across luminosities at a given redshift. Since we showed that the $z'$-band LF is a reasonable proxy for the SMF,  it is tempting to analyse the evolution of the rest-frame optical LF obtained in the present work in terms of the evolution of the dark-matter Halo Mass Function (HMF).

We performed a first analysis as follows.  We applied a simple abundance matching technique (\citealt{vale2004}) consisting in matching the cumulative number density of the LF to that of \citet{behroozi2013} HMF  obtained from \texttt{HMFcalc}\footnote{\url{http://hmf.icrar.org/} - we used the python implementation from \url{https://github.com/steven-murray/hmf}} \citep{murray2013} and recovered the evolution in mass of the halo mass function at $z\ge 5$ relative  to $z\sim4$. Given the rapid evolution of the HMF in this range of redshift, we adopted the HMFs at $z=3.78, 4.95, 5.76$ and $6.87$, corresponding to the (median) photometric redshifts of the stacked SEDs. The HMFs assumed $\sigma_8=0.81$. We found a relative displacement in halo mass of 0.76, 0.52 and 0.35~dex, corresponding to $\sim1.9$, $\sim1.3$ and $\sim0.9$~mag from $z\sim7, 6$ and $5$ to $z\sim4$, respectively. Figure \ref{fig:LF4HMF} shows the result of applying the above offsets to the $\mathcal{M}^*_{z'}$ of the $z\sim4$ Schechter parameterization.  The solid blue curve  marks the best-fit Schechter function at $z\sim4$, while the dashed curves   represent the $z\sim4$ LF rigidly shifted by the corresponding  amount at $z\ge5$. The agreement between the predicted and observed LF is very good at all redshifts, suggesting that the $z'$-band LF could trace the evolution of the HMF.

\subsection{Evolution of $M^*_h$}
\label{sect:mh}

\begin{figure*}
\includegraphics[width=17.5cm]{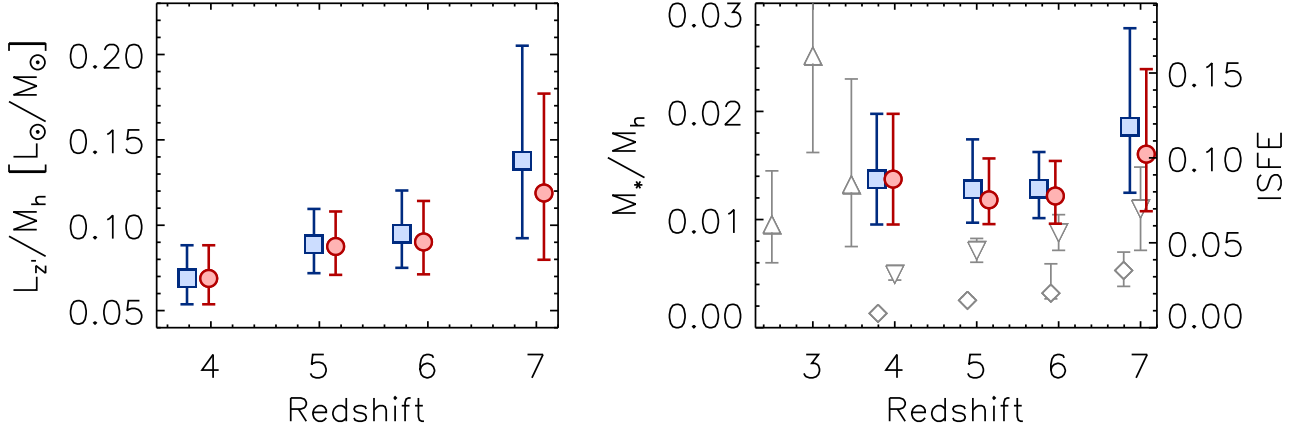}
\caption{{\bf Left panel:} Ratio between the luminosity and the halo mass from abundance matching at 1) constant cumulative number density (blue squares with errorbars) and 2) cumulative number density evolving following \citet[red circles with errorbars]{behroozi2013b}. The red points have been arbitrarily shifted by $\delta z=0.2$ to improve readability. The points show a mild indication of increase with redshift of the $L_{z'}/M_h$ ratio, although the large uncertainties make it also consistent with no evolution. {\bf Right panel:} Stellar-to-halo mass ratios (\citealt{chabrier2003} IMF), recovered from the values presented in the left panel, by applying the $M_*/L_{z'}$ relation in Sect. \ref{sect:ml_ratios} (same plotting conventions of the left panel). The grey points present stellar-to-halo mass ratios from the literature: \citet[open upward triangles]{durkalec2015}; \citet[open downward triangles, converted to \citealt{chabrier2003} IMF]{finkelstein2015b}  and \citet[open diamonds]{harikane2016}. We warn the reader that all these estimates refer to different $M_h$, making straight comparisons difficult to interpret (see text for more details). The y-axis on the right presents  the integrated star-formation efficiency (ISFE - also called the stellar baryon fraction), i.e., $M_*/M_h$ in units of $\Omega_b/\Omega_m$.  No clear evidence is found for an evolution of the ISFE with redshift. \label{fig:hmf_fix_cnd}}
\end{figure*}

In this section we discuss the evolution of those halos associated to a constant cumulative number density of $3.1\times10^{-5}$~Mpc$^{-3}$ over $4\lesssim z\lesssim 7$. This value corresponds to the cumulative number density of $\mathcal{M}^*_{z'}$ galaxies at $z\sim4$ ($M_{z'}=-23.38$~mag) and it allows us to recover the corresponding absolute magnitude up to $z\sim7$ with strongly reduced dependence (i.e., $\lesssim 0.5$~mag) on the extrapolation of the LFs to magnitudes fainter than actually observed. Table \ref{tab:hmf_fix_cnd} lists a compilation of the values of the main parameters recovered with our analysis.

\begin{table*}
\caption{Values of the main observables from our abundance matching analysis. \label{tab:hmf_fix_cnd}}
\begin{tabular}{ccccccccc}
\hline
\hline
& $z$ & Cum. Den.\tablenotemark{a} & $M_{z'}$ & $M_h$ & $L_{z'}/M_h$ & $M_*$ & $M_*/M_h$ & ISFE\tablenotemark{b} \\
& bin & $[\log(\mathrm{Mpc}^{-3})]$ & [mag] & $[\log(M_h/M_\odot)]$ & $[L_\odot/M_\odot]$ &  [$\log(M_*/M_\odot)$] & & \\
\hline
Fixed den. &  3.78 & $ -4.51$ & $-23.39\pm 0.27$ & $12.32$ & $0.069^{+0.019}_{-0.015}$ & $10.46\pm 0.16$ & $0.014^{+0.006}_{-0.004}$ & $0.087^{+0.038}_{-0.027}$ \\
 & 4.95 & $ -4.51$ & $-22.63\pm 0.23$ & $11.91$ & $0.089^{+0.021}_{-0.017}$ & $10.02\pm 0.13$ & $0.013^{+0.005}_{-0.003}$ & $0.081^{+0.029}_{-0.021}$ \\
 & 5.76 & $ -4.51$ & $-22.09\pm 0.26$ & $11.67$ & $0.095^{+0.025}_{-0.020}$ & $ 9.77\pm 0.10$ & $0.013^{+0.003}_{-0.003}$ & $0.082^{+0.022}_{-0.017}$ \\
 & 6.87 & $ -4.51$ & $-21.82\pm 0.43$ & $11.40$ & $0.138^{+0.067}_{-0.045}$ & $ 9.67\pm 0.17$ & $0.019^{+0.009}_{-0.006}$ & $0.118^{+0.058}_{-0.039}$ \\
 & & & & & & & & \\
 Evol. den. &  3.78 & $ -4.51$ & $-23.39\pm 0.27$ & $12.32$ & $0.069^{+0.019}_{-0.015}$ & $10.46\pm 0.16$ & $0.014^{+0.006}_{-0.004}$ & $0.087^{+0.038}_{-0.027}$ \\
 & 4.95 & $ -4.32$ & $-22.41\pm 0.23$ & $11.83$ & $0.088^{+0.020}_{-0.017}$ & $ 9.90\pm 0.12$ & $0.012^{+0.004}_{-0.002}$ & $0.075^{+0.024}_{-0.015}$ \\
 & 5.76 & $ -4.14$ & $-21.65\pm 0.26$ & $11.51$ & $0.090^{+0.024}_{-0.019}$ & $ 9.60\pm 0.10$ & $0.012^{+0.003}_{-0.003}$ & $0.078^{+0.021}_{-0.016}$ \\
 & 6.87 & $ -3.88$ & $-21.02\pm 0.43$ & $11.14$ & $0.119^{+0.058}_{-0.039}$ & $ 9.34\pm 0.17$ & $0.016^{+0.008}_{-0.005}$ & $0.102^{+0.050}_{-0.034}$ \\
 \hline
\end{tabular}
\tablenotemark{a}{Cumulative number density adopted for the abundance matching}
\tablenotetext{b}{Integrated Star-Formation Efficiency $\equiv (M_*/M_h)/(\Omega_b/\Omega_m)$, $\Omega_b/\Omega_m\equiv0.157$.}
\end{table*}

The abundance matching performed through cumulative number density implicitly assumes that each halo contains one and only one galaxy and that halos of the same mass contain galaxies of the same $z'$ luminosity ($\sim$stellar mass). Indeed, recent measurements have shown that the scatter between halo mass-stellar mass relation is quite small, $\sim0.15-0.20$~dex (e.g., \citealt{conroy2009, moster2010, tinker2016}, but see also \citet{gu2016} who found scatter of up to 0.32~dex).

The cumulative number densities of the LFs were computed adopting the Schechter parameterization presented in Sect. \ref{sect:lf_schechter}. For the LFs at $z\ge5$, we adopted the best-fit Schechter functions obtained when the characteristic magnitude was assumed to be the only free parameter of the fit,  coinciding with the case of pure luminosity evolution.

The values of the absolute magnitudes we obtain from our procedure are $M_{z'}\sim-23.39, -22.63, -22.09, -21.82$~mag for the $z\sim4$, $z\sim5$, $z\sim6$ and $z\sim7$ cases, respectively. These values are consistent within $1\sigma$ with the characteristic magnitudes of our Schechter fit. This is not surprising, considering the pure luminosity evolution of the Schechter fits themselves.

The matches to the cumulative number density performed on the HMFs resulted in halo masses $\log(M_h/M_\odot)=12.32, 11.91, 11.67, 11.40$ for the $z\sim4$, $z\sim5$, $z\sim6$ and $z\sim7$ cases, respectively. These displacements correspond to an evolution in the halo mass of $\mathcal{M}^*_{z'}$ galaxies of $\sim0.9, \sim0.6$, and $\sim0.3$~dex from $z\sim7, 6, 5$ to $z\sim4$, respectively. We note, however, that a rigid displacement in mass of the HMF is sufficient to reproduce the HMF evolution at these redshifts only for halo masses $\log(M_h/M_\odot)\gtrsim12$; at lower halo masses, the displacement in mass must be coupled to a steepening with redshift of the low-mass end slope. 

We can now use the above results on the evolution of the luminosity and of the halo mass to recover the evolution with redshift of the light-to-halo mass for galaxies at fixed cumulative number density.  Combining the two we obtain $L_{z'}/M_h\sim0.069, 0.089, 0.095, 0.138$ in units of $L_\odot/M_\odot$. These values are also presented in the left panel of Figure \ref{fig:hmf_fix_cnd} and suggest a mild increase with redshift (a factor $\lesssim 2\times$), although the large uncertainties make them consistent with a constant value across the 800~Myr of cosmic time from $z\sim4$ to $z\sim7$.

Using the results on the $M_*/L_{z'}$ from Sect. \ref{sect:ml_ratios}, we can convert the $L_{z'}/M_h$ into $M_*/M_h$. The result of this is shown in the right panel of Figure \ref{fig:hmf_fix_cnd}. The corresponding values are listed in Table \ref{tab:hmf_fix_cnd}. The stellar-to-halo mass ratio does not present any significant evolution with redshift. In the same panel we convert the $M_*/M_h$ into the integrated star-formation efficiency (ISFE i.e., $M_*/M_h$ in units of $\Omega_b/\Omega_m$ - \citealt{conroy2009}, and equivalent to the stellar baryon fraction - \citealt{finkelstein2015b}), using $\Omega_b/\Omega_m=0.157$ \citep{planck2015}. Our measurements are consistent with the ISFE being constant with redshift. We stress here that this result refers to $M_h\gtrsim 10^{12}M_\odot$ and it does not exclude the existence of evolution with redshift at lower halo masses. We defer a more complete analysis on the dependence of the $M_*/M_h$ ratio with halo mass to a future work. Furthermore, our samples at $z\sim5-7$ are entirely based on LBG selection. If non-negligible numbers of redder (dustier/more evolved) galaxies exist at these epochs, they would affect SMF (and likely its massive end, e.g. \citealt{stefanon2015, caputi2015}) and, consequently, the recovered $M_h$.

Matching galaxies at a constant cumulative number density, however, does not consider the effect of major mergers in the galaxy ranking. We therefore repeated the same analysis using a cumulative number density evolving with redshift following the recipe of \citet{behroozi2013b}. The results are listed in Table \ref{tab:hmf_fix_cnd}, and plotted as red circles in Figure \ref{fig:hmf_fix_cnd}. No significant difference with the constant cumulative number density match is observed.  We also note that the values for the $z\sim7$ bin rely on the extrapolation of the LF to luminosities below those currently probed by our sample. Those measurements should then be treated with caution. 

In the right panel of Figure \ref{fig:hmf_fix_cnd} we also plot recent estimates of the $M_*/M_h$ from the literature: \citet{durkalec2015}, \citet[converted to a \citet{chabrier2003} IMF by applying a factor 0.55]{finkelstein2015b} and \citet{harikane2016}.   \citet{durkalec2015} applied the measurements of the two-point correlation function to a halo occupation model to recover the halo mass of samples of galaxies at $z\sim2-5$ with  spectroscopic redshift from the VIMOS Ultra Deep Survey (VUDS - \citealt{lefevre2015}). \citet{finkelstein2015b} measured the evolution of the $M_*/M_h$ from abundance matching the $z\sim4-7$ UV LF. \citet{harikane2016} recovered $M_*/M_h$ from the clustering of LBGs selected at $z\sim4-7$ from a variety of programs, including CANDELS, the \textit{Hubble} Frontier Fields (PI: J. Lotz) and Subaru Hyper-Suprime Cam Subaru Strategic Program (PI: S. Miyazaki).

At face value, our measurements are consistent with those of \citet{durkalec2015} at $z\sim4$ and with those of \citet{finkelstein2015b} at $z\sim6-7$; however, they are inconsistent with those of \citet{harikane2016} over the full range of redshift, and with those of \citet{finkelstein2015b} at $z\sim4-5$.  Recently \citet{mancuso2016}, applying abundance matching to the evolution of the SFR function recovered from UV+far-IR data,  found indication for a non-evolving $M_*/M_h$ ratio at $z\ge4$.

The $M_*/M_h$ measurements presented in the right panel of Figure \ref{fig:hmf_fix_cnd} were obtained from a variety of methods, and ultimately refer to different $M_h$ estimates, making the straight comparison difficult to interpret. Specifically, our measurements of the halo mass are based on a constant cumulative number density match with $\log(M_h/M_\odot)\sim12.3, 11.9, 11.7, 11.4$ at $z\sim4,5,6,7$, respectively. The increase with redshift of the $M_*/M_h$ of \citet{harikane2016}  refers to a \emph{fixed} $\log(M_h/M_\odot)=11$ across $z\sim4-7$. \citet{finkelstein2015b} report an increase with redshift of the ISFE $\propto (0.024\pm0.07)\times z$.  However, this trend is most likely driven by the point at $z\sim4$, and originated by limited evolution of the UV LF between $z\sim4$ and $z\sim5$ ($\Delta\mathcal{M}_\mathrm{UV}^*\sim0.08\pm0.16$~mag) observed by \citet{finkelstein2015a}, in contrast to the large luminosity evolution ($\Delta\mathcal{M}_{z'}^*\sim0.76\pm0.22$~mag) observed in our work over the same redshift interval. This, together with the constant characteristic magnitude of the UV LF (i.e., $\approx$ constant SFR) assumed as criterion for the abundance matching, generates a reference cumulative number density decreasing with redshift, and halo masses $\log(M_h/M_\odot)\sim11.9, 11.7, 11.6, 11.3$. We note here that our measurements are consistent with those of \citet{finkelstein2015b} at $z\sim6-7$, i.e., where the $M_h$ recovered by the two teams are more similar, while they are inconsistent at $z\sim4-5$, where the $M_h$ differ. Finally \citet{durkalec2015} estimates refer to a diversity of halo masses and redshift ranges: $\log(M_h/M_\odot)\sim11.1, 11.5, 11.2$ for $z_\mathrm{mean}\sim2.5, 3.0, 3.5$, respectively.

Our finding of $M_*/M_h$ independent of redshift is also qualitatively in agreement with recent estimates of the galaxy bias, observed to be nearly constant over $z\sim4-6$ (\citealt{barone2014}), although a measurement at $z\sim7$ from the same work seems to suggest a potential change of the star formation efficiency at earlier epochs.

The picture is not settled even from a theoretical perspective. Indeed, some of the models predict an \emph{increase} with cosmic time of the $M_*/M_h$ ratio for a fixed $M_h$ (e.g., \citealt{somerville2015}). Other models find that $M_*/M_h$  decreases with cosmic time at fixed halo mass (\citealt{moster2013, behroozi2015}). However, when considering the evolution of the same population of galaxies (through an evolving number density), \citet{behroozi2015} found that $M_*/M_h$ is nearly independent on redshift. Finally, other models, instead, result in $M_*/M_h$ to be insensitive to redshift (e.g., \citealt{oshea2015,mutch2016}). Some semi-empirical models have been able to reproduce the evolution of the UV LF from $z\sim2$ to $z\sim10$ under the assumption that, for star-forming galaxies, the $M_*/M_h$ depended on $M_h$ but not on redshift (\citealt{trenti2010, tacchella2013, mason2015}).

\section{Summary and Conclusions}

The main aim of this work was to measure the rest-frame $z'$-band luminosity function (LF) of field galaxies and to study its evolution  at $z\ge4$. The rest-frame $z'$ band was selected for three reasons: 1) it is not contaminated by strong emission from nebular lines; 2) light in this wavelength range is dominated by lower mass, long-living stars; and 3) it can be probed up to $z\sim8$ using the current \emph{Spitzer}/IRAC data.  These characteristics suggest it can provide a complementary basis for dealing with stellar mass measurements at high redshift, minimizing the potential systematic effects that can affect stellar mass measurements.

We therefore assembled samples of Lyman Break Galaxies (LBGs) at $z\sim4$, $z\sim5$, $z\sim6$ and $z\sim7$, selected over the GOODS-N and GOODS-S fields. The $z\sim4$ sample was complemented by galaxies with photometric redshifts $3.5<z_\mathrm{phot}<4.5$ extracted from a 37-band far-UV-to-$8.0\mu$m $K_\mathrm{s}$-detected photometric catalog based on UltraVISTA DR2. The larger $z\sim4$ co-moving volume provided by the UltraVISTA data allowed us to gain statistics on the rarer more luminous and/or redder galaxies.

The GOODS-N/S sample takes advantage from the recently released full depth IRAC maps \citep{labbe2015}, obtained from the combination of all the IRAC programs carried out so far over these fields, namely IGOODS, IUDF, GOODS, ERS, S-CANDELS, SEDS and UFD2.  These maps reach a depth of $\sim25.8$ mag and $\sim24.5$~mag in the $4.5\mu$m and $5.8\mu$m bands respectively ($2\farcs0$ diameter aperture, $5\sigma$), although the coverage is highly inhomogeneous. Similarly, the UltraVISTA catalog benefits from IRAC $3.6\mu$m and $4.5\mu$m mosaics which combine the S-COSMOS, S-CANDELS and SPLASH programs and reach a depth of $\sim22.5$~mag ($2\farcs0$ diameter aperture, $5\sigma$).

We further selected our sample based on the S/N in the IRAC band or bands closer to the rest-frame $z'$ band. Specifically, the final $z\sim4$ sample was selected to have $S/N>5$ in the $4.5\mu$m while the $5<z<7$ samples were selected to have $S/N>4$ in the inverse-variance weighted combination of S/N in the $5.8\mu$m and $8.0\mu$m bands. Our final composite sample included 2098, 72, 10 and 2 objects, for the $z\sim4, 5, 6$ and 7 redshift bins, respectively. Although the $z'$ band is covered by the $5.8\mu$m and $8.0\mu$m IRAC data up to $z\sim8$, we do not register any LBG galaxy at $z\sim8$  which also satisfies our selection criteria on the S/N of IRAC fluxes.

Our main results are as follows:

\begin{enumerate}
\item At $z\sim4$ and for absolute magnitudes $M_{z'}$ fainter than $\sim-23$~AB, galaxies follow a linear relation on the $M_\mathrm{UV}-M_{z'}$ plane, with slope $\sim0.8$. This correlation breaks at  $M_{z'}\lesssim-23$~AB: the $M_\mathrm{UV}$ of these galaxies covers the full range of values observed for galaxies with fainter $M_{z'}$ (Figure~\ref{fig:uvopt}).
\item We performed stacking analysis and measured the $M_*/L_{z'}$ of galaxies segregated according to their redshift and absolute magnitude $M_{z'}$. The $M_*/L_{z'}$ at $z\sim4$ is independent of $M_{z'}$ for $M_{z'}\gtrsim-22.5$; at brighter $M_{z'}$ the $M_*/L_{z'}$ increases with luminosity following a power-law. The  $M_*/L_{z'}$ at $z\ge5$ are consistent with those observed at $z\sim4$, although the associated large uncertainties may hide a different behaviour (Figure~\ref{fig:stack} and \ref{fig:ML}).
\item We computed the LF in the rest-frame $z'$ band, using the $V_\mathrm{max}$ estimator, in four different redshift bins: $z\sim4$, $z\sim5$, $z\sim6$ and $z\sim7$. We admit freely that our single bin measurement at $z\sim7$ does not allow us to set stringent limits on the shape of the $z\sim7$ LF. The LF shows evolution from $z\sim7$ to $z\sim4$. Schechter fits to the $V_\mathrm{max}$ LF marginally prefer pure evolution in luminosity over a pure evolution in density (Figure~\ref{fig:LF}). 
\item The non-decreasing  $M_*/L_{z'}$ with luminosity (corresponding to an injective mapping) allowed us to apply a simple conversion from luminosity to stellar mass. We therefore converted our LF measurements into SMF using the $M_*/L_{z'}$ recovered from the stacking analysis at $z\sim4$. The obtained SMFs are consistent with the average SMF determination from the literature. Despite the relaxed S/N cuts in IRAC flux applied to our samples, the lower stellar mass over which we recover our SMFs is $\sim5-10\times$ larger than typical lower limits from the literature (Figure~\ref{fig:SMF}).
\item Evolution in the halo mass relative to $z\sim4$ recovered from abundance matching the halo mass functions reproduces the luminosity evolution of the LF at $z\gtrsim4$ (Figure \ref{fig:LF4HMF}). The stellar-to-halo mass ratio at fixed cumulative number density shows no strong evidence for evolution with redshift over $4<z<7$ (Figure \ref{fig:hmf_fix_cnd}).
\end{enumerate}

The above results allow us to draw the following conclusions:

\begin{enumerate}
\item The current depth of \emph{Spitzer}/IRAC data revealed to be sufficient to probe the regimes in rest-frame UV luminosities both where the rest-UV luminosities are correlated with stellar mass and where they are not.
\item The existence at  $z\sim4$ and $z\sim5$ of LBGs luminous in the rest-frame $z'$ band and spanning a broad range in UV luminosities  suggests that the adoption of the UV LF for SMF estimates may be affected by systematics. Specifically, samples of galaxies selected to have a narrow range in UV luminosities potentially include a combination of high and low mass objects, and ultimately can introduce an over-estimate of the low-mass end slope and an underestimate of the densities at the high-mass end.
\item The higher values of the lower stellar mass bin in our SMFs compared to recent determination from the literature, arising from the S/N cuts applied to the IRAC fluxes,  suggests that current low-mass end of the SMFs at $z\gtrsim4$ might be based on low S/N flux measurements ($\sim1-2\sigma$ upper limits) in the observed IRAC bands most sensitive to the stellar mass (i.e., $5.8\mu$m and $8.0\mu$m). Higher S/N measurements are available from $3.6\mu$m and $4.5\mu$m data. However these bands at $z\gtrsim5$ are contaminated by nebular emission which can potentially bias the stellar mass estimates, given our still limited knowledge on the emission line intensities of high-redshift star forming galaxies.
\item The rest-frame $z'$ band LF can be a valid proxy for SMFs and HMF measurements at $z\gtrsim4$, and complementary to SMF estimates based on individual stellar mass measurements. The nearly flat dependence of the $M_*/L_{z'}$ on $M_{z'}$ increases this confidence.

\end{enumerate}

This work is largely based on data from the cryogenic programs of \emph{Spitzer}/IRAC. While the depth of the $3.6\mu$m- and $4.5\mu$m- band data can still be improved through non-cryogenic programs, the sensitivity of JWST/MIRI provides the only opportunity for increasing the depth at wavelengths $\lambda>5\mu$m, necessary for improving current estimates of stellar masses at $z\gtrsim5$.\\

\acknowledgments

We are appreciative to Adriano Fontana for a very helpful referee report which greatly improved this paper. MS would like to thank Adriano Fontana for constructive discussions. MS and RB are grateful to Karina Caputi, Yuichi Harikane, Michele Trenti and Stuart Wyithe for helpful feedback on an advanced draft of this manuscript. This work is based on data products from observations made with ESO Telescopes at the La Silla Paranal Observatory under ESO programme ID 179.A-2005 and on data products produced by TERAPIX and the Cambridge Astronomy Survey Unit on behalf of the UltraVISTA consortium. This work is based on observations taken by the CANDELS Multi-Cycle Treasury Program with the NASA/ESA HST, which is operated by the Association of Universities for Research in Astronomy, Inc., under NASA contract NAS5-26555. This work is based on observations taken by the 3D-HST Treasury Program (GO 12177 and 12328) with the NASA/ESA HST, which is operated by the Association of Universities for Research in Astronomy, Inc., under NASA contract NAS5-26555. 

\appendix

\section{Sample selection criteria}
\label{app:sel}

After applying the S/N cuts described in Sect.~\ref{sect:sample_assembly}, we further cleaned our sample, excluding those objects satisfying any of the following conditions: 1) the contribution to the $5.8\mu$m and $8.0\mu$m flux from neighbouring objects is excessively high; 2) the source morphology is very uncertain or confused making IRAC photometry undetermined; 3) the source is detected at X-rays wavelengths, suggesting it is a lower redshift AGN; 4) the source is at higher redshift, but its SED is dominated by AGN light; 5) LBGs with a likely $z<3.5$ solution from photometric redshift analysis. In the following paragraphs we will describe in more detail the above criteria and their effects on the sample size.

The broad IRAC PSF together with the unprecedented photometric depth of the IRAC mosaics in the GOODS-N, GOODS-S and UltraVISTA  fields can result in flux measurements potentially affected by contamination from brighter nearby objects. The procedure we adopted for the flux measurements  already deals with this problem by cleaning each source from its neighbours before performing the photometry. However, in some cases the flux at the position of the object of interest mostly comes from the bright neighbours, resulting in potentially very uncertain flux measurements. We therefore opted to further clean our sample by applying a cut on the maximum fraction of flux from neighbours contributing to the flux of each object before the neighbour-cleaning process. Specifically we excluded from our sample those sources whose neighbours were contributing more than 65\% to the total flux at the position of each source in our sample.  We also visually inspected the cutouts from the IRAC photometry and further excluded those sources showing a residual contamination from bright nearby sources.  In this step we removed  280 galaxies (211/69, for GOODS-N/S and UltraVISTA, respectively; of the 211 GOODS galaxies, 3 were at $z\sim6$ and 1 at $z\sim7$). In Section~\ref{sect:completeness} we describe the Monte Carlo simulation we implemented to evaluate the selection effects introduced by the above selection criteria in a statistical way.

We visually inspected the cutouts of the GOODS-N/S sample in the WFC3/$H_{160}$ band and of the UltraVISTA sample in the ACS/F814W, and excluded those objects with doubtful morphology, as e.g., it was the result of two or more distinct objects or the deblending from SExtractor was deemed inconsistent. Furthermore, the visual inspection also allowed us to identify and exclude objects with point-source morphology as either potentially AGN dominated or brown dwarfs contaminants. This was particularly important for the UltraVISTA sample, since the $K_\mathrm{s}$-band detection image is characterized by a  PSF FWHM $\sim 0\farcs8$, much broader compared to that of ACS or WFC3 (FWHM $\sim 0\farcs12$ and $\sim0\farcs2$, respectively).   This class of objects are subject to very inaccurate photometry, redshift classification and/or luminosity measurement. Through the above criteria we excluded 53 objects from the GOODS-N/S sample (1 at $z\sim7$) and 63 objects from the UltraVISTA sample.

Successively, we crossmatched our sample to catalogs of X-ray sources in the GOODS-N/S \citep{alexander2003, xue2011} and COSMOS fields \citep{cappelluti2009, elvis2009, paris2012}, and excluded all the matching sources as these are potential lower-redshift AGN contaminants. We identified 34 sources with an X-ray counterpart matching our initial sample, most of which at $z\sim4$ (29), and 4 at $z\sim5$. Furthermore we visually inspected all the observed SEDs to exclude either objects with very red, power-law like rest-frame optical/NIR slopes which could be signature of Type-1 AGN. In this step we flagged and removed from the sample a total of 74 sources (20/54).

Finally, we run EAzY \citep{brammer2008} on the sample of LBGs, and excluded those galaxies with $z_\mathrm{peak}<3.5$, sources with prominent secondary lower-$z$ solution or inconsistent SED. Through this step we excluded 109 sources (4 of which at $z\sim6$).

The final sample consists of 2098 galaxies at $z\sim4$ (1680 from the LBG sample and 418 from the UltraVISTA sample), 72 at $z\sim5$, 10 at $z\sim6$ and 2 objects at $z\sim7$.

\section{SEDs of the $z\sim5, 6$ and $7$ samples}

\label{app:SEDs}

\begin{figure*}
\includegraphics[width=17.5cm]{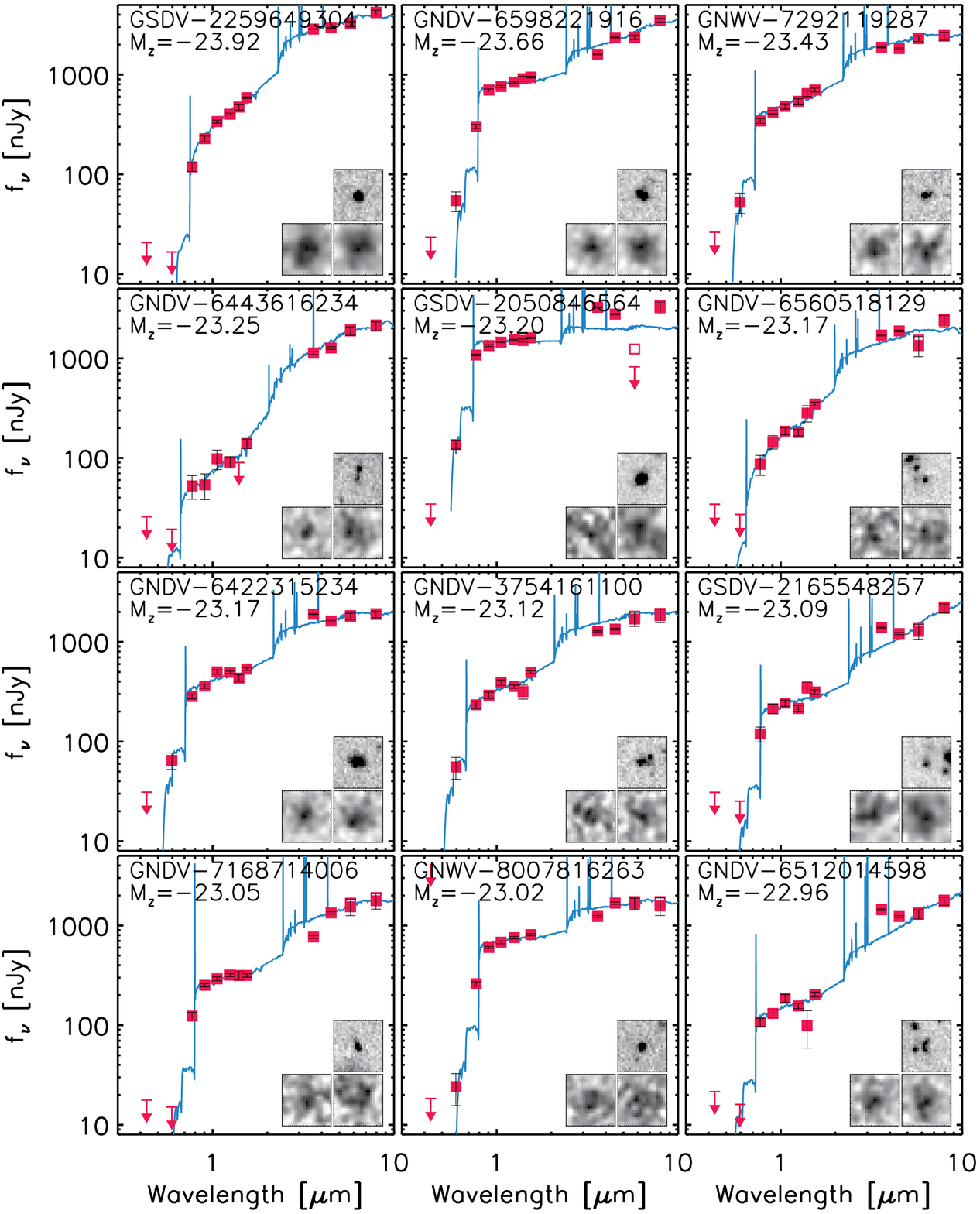}
\caption{SEDs of the 12 most luminous galaxies in the $z\sim5$ sample. Photometric measurements are marked by the solid boxed with errorbars, with arrows for $2\sigma$ upperlimits. The measurements for the $5.8\mu$m and $8.0\mu$m bands include the correction for the flux boosting. The original measurements for these two bands are shown as open boxes. The blue curve identifies the best-fit SED template from EAzY. In each panel, the three insets present the cutouts ($\sim3\farcs0$ side) in the WFC3 $H_{160}$ band (upper box) and neighbour-subtracted $5.8\mu$m and $8.0\mu$m bands (lower boxes, left and right, respectively). Labeled in the top left corner is the absolute magnitude $M_{z'}$.\label{fig:seds_z5}}
\end{figure*}

\begin{figure*}
\includegraphics[width=17.5cm]{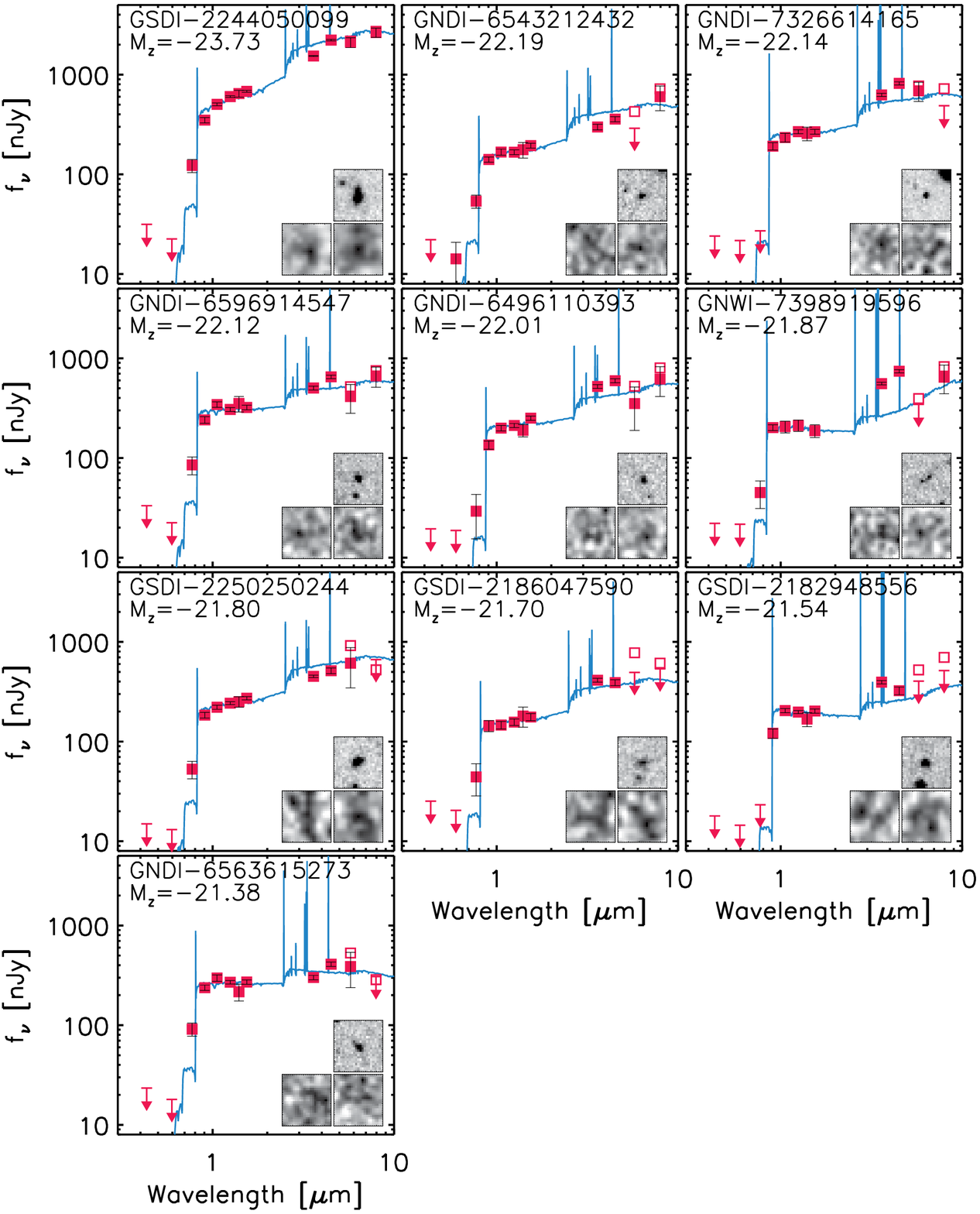}
\caption{SEDs of the $z\sim6$ sample. Other plotting conventions as for Figure \ref{fig:seds_z5} \label{fig:seds_z6}}
\end{figure*}

\begin{figure}
\includegraphics[width=17.5cm, trim=0 500 0 0]{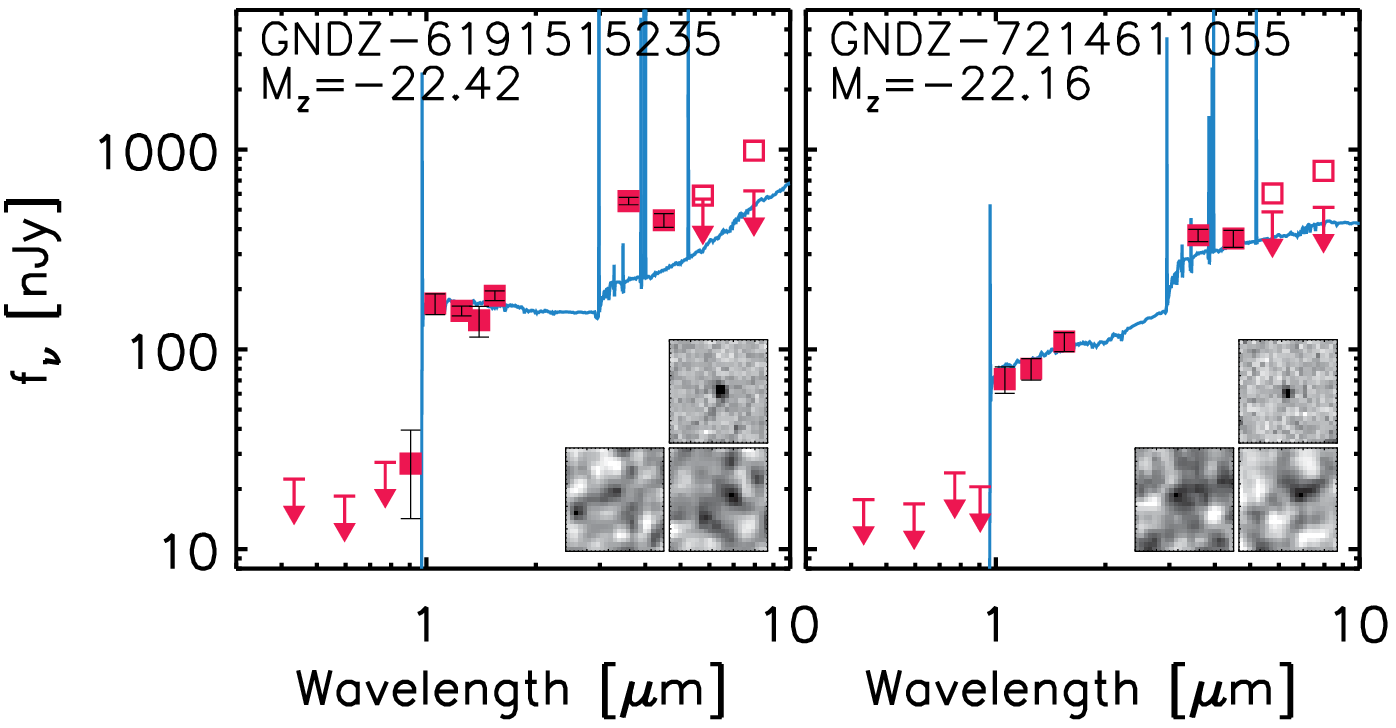}
\caption{SEDs of the  galaxies at $z\sim7$. Other plotting conventions as for Figure \ref{fig:seds_z5} \label{fig:seds_z7}}
\end{figure}

In Figure \ref{fig:seds_z5} we present the SEDs of the 12 most luminous galaxies included in the $z\sim5$ sample, while Figures \ref{fig:seds_z6} and \ref{fig:seds_z7} show the full sample of galaxies at $z\sim6$ and $z\sim7$, respectively.\\

Most of the galaxies are characterised by small, compact sizes in the $H_{160}$. Noticeably, GSDI-2244050099, at $z\sim6$, is among the most luminous galaxies of the full sample, including $z\sim4$. Its apparent size is larger than the average size of the galaxies in the $z\sim5, 6$ and $7$ samples. Careful inspection of the WFC3/$H_{160}$ cutout does not show any indication of clumpiness. However, in ACS/F184W we observe 2 possible components, separated by $\sim0\farcs45$ ($\sim2.6$~kpc at $z=6$). We opted for including it in our sample given the small separation between the two components visible only in the ACS data, the consistency of the SED and the fact that the IRAC flux appears to be centered at the position of the brighter component. This object constitutes the unique element of the highest luminosity bin of the $z\sim6$ LF. Noteworthy, even assuming a factor 2 overestimate of the IRAC flux, the resulting absolute magnitude would still be consistent with the highest luminosity bin.

\end{document}